\documentclass[a4paper,11pt]{article}

\usepackage{amsmath,amssymb,amstext,amsfonts}
\usepackage{amsthm, mathrsfs, thmtools}
\usepackage{graphicx,subfigure}
\usepackage{url}
\usepackage{authblk}
\usepackage[round,colon]{natbib}
\usepackage{dsfont} 
\usepackage{multirow} 
\usepackage{sectsty} 
\usepackage{setspace}
\usepackage{lifecon}
\usepackage{enumitem}

\usepackage[hmargin=1in,vmargin=1.5in]{geometry}

\usepackage[title]{appendix}

\usepackage[format=hang,font=small,labelfont=bf]{caption} 

\subsectionfont{\mdseries\normalfont} 

\subsubsectionfont{\mdseries\itshape} 

\newcommand{\indicatorfn}{\mathds{1}}

\newcommand{\naturalnumbers}{\mathbb{N}}

\newcommand{\expectation}{\mathbb{E}}

\newcommand{\numalive}{L}
\newcommand{\numdead}{N}

\newcommand{\fnbobsurvives}{f}
\newcommand{\fnbobeither}{g}
\newcommand{\fnsummand}{h}

\newcommand{\notfundA}{F^{\textrm{A}}}

\newcommand{\consumBzero}{c_{0}^{\textrm{B}}}
\newcommand{\consumBone}{c_{1}^{\textrm{B}}}

\newcommand{\consumXzero}{c_{0}^{\textrm{X}}}
\newcommand{\consumXone}{c_{1}^{\textrm{X}}}

\newcommand{\notfundB}{F^{\textrm{B}}}
\newcommand{\notfundX}{F^{\textrm{X}}}

\newcommand{\probsurvive}{p}
\newcommand{\probdie}{q}

\newcommand{\probsurviveA}{p^{\textrm{A}}}
\newcommand{\probdieA}{q^{\textrm{A}}}

\newcommand{\probsurviveB}{p^{\textrm{B}}}
\newcommand{\probdieB}{q^{\textrm{B}}}

\newcommand{\probsurviveX}{p^{\textrm{X}}}

\newcommand{\fundfraction}{\textrm{MEA}}

\newcommand{\wealthrich}{\beta}

\theoremstyle{plain}

\theoremstyle{definition}

\theoremstyle{remark}

\newcommand{\grpXchars}{\mathbf{z}^{X}}
\newcommand{\grpAchars}{\mathbf{z}^{A}}
\newcommand{\grpBchars}{\mathbf{z}^{B}}

\newcommand{\GSA}{group self-annuitization scheme}
\newcommand{\GSAspace}{group self-annuitization scheme }

\newcommand{\GSAabbrev}{GSA}
\newcommand{\GSAabbrevspace}{GSA }

\newcommand{\PAFspace}{pooled annuity fund }

\newcommand{\PAFabbrev}{PAF}
\newcommand{\PAFabbrevspace}{PAF }

\newcommand{\AOFabbrev}{AOF}
\newcommand{\AOFabbrevspace}{AOF }

\newcommand{\namembrB}{Bob}
\newcommand{\namembrBspace}{Bob }

\newcommand{\AOFspace}{annuity overlay fund }

\title{Actuarial fairness and solidarity in pooled annuity funds}

\author{Catherine Donnelly\thanks{Department of Actuarial Mathematics and Statistics, and the Maxwell Institute for Mathematical Sciences, Heriot-Watt University, Edinburgh EH14 4AS, United Kingdom. Email: {\tt C.Donnelly@hw.ac.uk}.  Tel: +44 131 451 3251.  Fax: +44 131 451 3249.}}

\begin{document}

\maketitle

\begin{abstract}
Various types of structures that enable a group of individuals to pool their mortality risk have been proposed in the literature.  Collectively, the structures are called pooled annuity funds.  Since the pooled annuity funds propose different methods of pooling mortality risk, we investigate the connections between them and find that they are genuinely different for a finite heterogeneous membership profile.

We discuss the importance of actuarial fairness, defined as the expected benefits equalling the contributions for each member, in the context of pooling mortality risk and comment on whether actuarial unfairness can be seen as solidarity between members.  We show that, with a finite number of members in the fund, the group self-annuitization scheme is not actuarially fair: some members subsidize the other members.  The implication is that the members who are subsidizing the others may obtain a higher expected benefit by joining a fund with a more favourable membership profile.  However, we find that the subsidies are financially significant only for very small or highly heterogeneous membership profiles.
\end{abstract}

\hspace*{2cm}

{ \bf Keywords:} pensions; group self-annuitization; annuity overlay; mutual risk-sharing; mortality risk. \\

\pagestyle{myheadings}
\thispagestyle{empty}
\markboth{}{Actuarial fairness and solidarity in pooled annuity funds}

\newpage

\section{Introduction}
As more and more people are encouraged to save for their retirement, it is important to design stable, robust and transparent pension schemes that allow them to pool risks together.  For example, at the moment in the U.K, individuals not in a defined-benefit pension scheme have, in the main, one of two options at retirement: buy a life annuity or drawdown their assets.  There should be alternatives for them that allow them to benefit from pooling mortality risk\footnote{We define mortality risk as the knowledge of the probability distribution of someone's future lifetime but not the exact future lifetime of that individual.  We distinguish this from longevity risk, which is the risk arising from not knowing the probability distribution of someone's future lifetime.  Mortality risk can be eliminated by pooling a sufficient number of lives whereas longevity risk cannot.} without having to buy the lifetime income guarantee offered by a life insurance company.

A number of alternative pension schemes which allow people to pool mortality risk have been proposed in the literature and we examine three of them.  They each propose, either explicitly or implicitly, a rule for sharing mortality risk among the members of the fund.  In these pooled annuity funds, participants in the schemes bear the entirety of any investment, mortality and longevity risk.  There is no sponsoring employer or life insurance company to which members can appeal for additional funds.

Actuarial fairness is a very important concept in such pooled annuity funds.  If a pooled annuity fund is actuarially fair then the present value of the contributions paid by a member should equal the present value of the benefits received by that member.  The benefits may be expressed as consumption payments or they may be the member's fund value at a future point in time, depending on the type of pooled annuity fund under consideration.

The reason why actuarial fairness is important is that there is no insurance aspect in any of the pooled annuity funds.  They are mutual risk-sharing arrangements and there is no guarantee as to the amount of the benefits paid by the fund.  If a fund is actuarially unfair, then one or more members are subsidizing others in the fund.  It means that some people in the fund can expect, in a probabilistic sense, to receive less than what they have contributed to the fund.  Correspondingly, others in the same fund can expect to receive more.

Unsurprisingly, we find that, with a large enough number of members so that the mortality risk is fully diversified, and assuming that there is no longevity risk, the pooled annuity fund structures are all actuarially fair.  However, with a finite number of members, it is no longer true.   This is important since we do not know how these funds could be constructed in practice.  For example, defined-benefit pension schemes are structures that pool mortality risk.  Yet in the U.K. there are over 2\,000 defined-benefit pension schemes with fewer than 100 members \citep{PPF2013.misc}.  The inability of such small schemes, which can be highly heterogeneous, to adequately pool mortality risk is rarely, in the author's experience, given sufficient consideration (see \citealt{donnelly11.article} for an analysis of the potential impact of inadequately pooled mortality risk in defined-benefit pension schemes).  The financial consequences for the individual member of sharing mortality risk in the pooled annuity funds should be examined, analysed and communicated to them.

We could interpret the lack of actuarial fairness as solidarity between members.  The rule for sharing the mortality risk is fixed in advance.  Yet in an actuarially unfair fund, the solidarity is only one way.  The fund's members are not offering genuine mutual support to each other: we know in advance which members are subsidizing (in expectation) the others.  So it seems that we should not re-label actuarial unfairness as solidarity as there is no uncertainty about who bears the expected losses due to the actuarial unfairness.

Instead, solidarity occurs since actuarial fairness is defined as ``fairness in expectation''.  In some future states of the world, a member may lose financially from pooling mortality risk.  In other future states of the world, the same member may gain financially.  The solidarity occurs from the willingness of the fund members to accept whatever gain or loss befalls them individually, without knowing in advance which it will be.

As we define it in this paper, actuarial fairness means that each member can expect to get back from the fund the same amount as what they have contributed or put at risk.  It is a concept of fairness between the members of the fund: one member should not benefit at the expense of another.  There are other possible definitions of fairness.  In \citet[pages 504--505]{piggottetal05.article}, the concept of actuarial fairness is predicated on the perfect pooling of mortality risk: the calculation of future consumption payments assumes that mortality risk is fully diversified regardless of the true scheme membership profile.  Although consumption payments are adjusted for the deviations of the observed scheme mortality experience from the perfectly pooled situation, the interpretation remains that fairness in this alternative definition is relative to a scheme in which mortality risk is fully diversified.

We consider the following mortality risk-sharing schemes:
\begin{itemize}
	\item[\GSAabbrev] The \GSAspace of \citet{piggottetal05.article}, which is proposed for any heterogeneous group of members. 
	\item[\PAFabbrev] The \PAFspace of \citet{stamos08.article}.  This fund is proposed for homogeneous groups only, in the sense that each member invests the same amount of wealth, consumes the same amount, has the same risk preferences and the random variables modelling future lifetimes are independent and identically distributed.
	\item[\AOFabbrev] The \AOFspace of \citet{donnellyetal14.article}, which is proposed for any heterogeneous group of members.  Assuming that the random variables modelling future lifetimes are independent and identically distributed, \citet{donnellyetal14.article} show that the fund is actuarially fair at each instant in time.  Consequently, it is actuarially fair over each member's lifetime.
\end{itemize}
 
Broadly, these schemes have similar aims: they are mechanisms to share mortality risk among groups of individuals.  So how different can they be?  The paper has two motivations:
\begin{itemize}
	\item To determine the connections between the above schemes.
	\item To determine if the schemes are actuarially fair.  This question has been answered only for the annuity overlay fund in \citet{donnellyetal14.article}.   If not, then there must be cross-subsidies between members.  Are these financially significant cross-subsidies?
\end{itemize}

Under the assumption of no longevity risk, we show the following.
\begin{itemize}
	\item For a homogeneous group, the \PAFspace of \citet{stamos08.article} with a specific choice of the consumption rate is equivalent to the \GSAspace of \citet{piggottetal05.article}.  While we suspect that this is already known (for example, see \citealt{qiaosherris12.article}), we have not seen a clear statement.
	\item For a homogeneous group, the \PAFspace of \citet{stamos08.article}, and hence the \GSAspace of \citet{piggottetal05.article}, is actuarially fair.
	\item For a heterogeneous group with a finite number of members, the \GSAspace of \citet{piggottetal05.article} is not actuarially fair.  The poorer members of the scheme benefit from subsidies from the richer members, and the young benefit from subsidies from the old (see also the numerical example in \citealt[Appendix I]{Sabin10.unpublished}).  However, the subsidies are only financially significant for very small or highly heterogeneous membership profiles. 
\end{itemize}

Another method of mortality risk-sharing that we do not consider in this paper is \citet{Sabin10.unpublished}, who proposes conditions to be satisfied by a mortality risk-sharing rule that distributes the wealth of the deceased among the survivors of the fund in order that it is actuarially fair.

\section{Connection of the \GSAabbrevspace with the \PAFabbrev}

In the \GSAspace (\GSAabbrev) introduced by \citet{piggottetal05.article}, the basic idea is that the consumption rate is calculated on an expected investment and mortality basis.  However, the consumption rate changes over time in line with the actual investment and mortality experience of the group.

As we see next, the \PAFspace (\PAFabbrev) of \citet{stamos08.article} is identical to the \GSAabbrev, for a specific choice of the consumption rate in the \PAFabbrevspace and a homogeneous group of people.  Note that the \PAFabbrevspace is defined only for a homogeneous group of people, so we are unable to compare it to the \GSAabbrevspace for a heterogeneous group.

\subsection{The homogeneous group} \label{SUBSEChomogroup}

Suppose there is a group of $\numalive$ individuals alive at time 0, with time measured in years.  These individuals are independent and identical copies of one another.  For example, they have the same future lifetime distribution and the same risk preferences.  Each individual has initial wealth 1 unit.

Let $\ddot{a}_{n}$ denote the expected present value at integer time $n$, of a life annuity payment of 1 unit per annum, paid at the start of each year beginning at time $n$ and calculated on a fixed investment and mortality basis.  We set investment returns equal to zero for simplicity.

\subsection{The \PAFabbrev} \label{SUBSECpaf}

Suppose that the homogeneous group join a \PAFabbrev, as described in \citet{stamos08.article}.  Then at time 0 the wealth of each member, before any consumption, is
\[
F_{0}^{\textrm{\PAFabbrev}} = 1.
\]
As members in the group die, their wealth is shared equally among the survivors.  For example, upon the first death in the group, each survivor's wealth increases from 1 unit to
\[
1 + \frac{1}{\numalive - 1}.
\]
Denote the number of deaths that have occurred by the end of the first year by $\numdead$.  Then the wealth of a survivor in the fund accumulates to
\[
F_{1}^{\textrm{\PAFabbrev}} = \left( 1 + \frac{1}{\numalive - 1} \right) \left( 1 + \frac{1}{\numalive - 2} \right) \cdots \left( 1 + \frac{1}{\numalive - \numdead} \right) = 1 + \frac{\numdead}{\numalive - \numdead}
\]
at time 1.

In the \PAFabbrev, members can withdraw money at any time but they must withdraw the same amount at the same time.  In other words, it is a group decision how much to withdraw from the fund at each point in time.

Assume each member consumes the amount $c_{0}^{\textrm{\PAFabbrev}}$ at time 0 and nothing else until time 1.  The remaining wealth of each member at time 0 is
\[
1 - c_{0}^{\textrm{\PAFabbrev}}.
\]
At time 1, the wealth of a surviving member is
\begin{equation} \label{EQNpafnotfundvalue}
F_{1}^{\textrm{\PAFabbrev}} = \left( 1 + \frac{\numdead}{\numalive - \numdead} \right) ( 1 - c_{0}^{\textrm{\PAFabbrev}} )
\end{equation}
before any consumption at time 1.

\subsection{The homogeneous \GSAabbrev}
Suppose instead that the homogeneous group join a \GSAabbrev, as described in \citet[Section ``A Simple Actuarial Analysis of GSA Plans'']{piggottetal05.article}.  The description of the \GSAabbrevspace focuses on the consumption of the participants.  However, by notionally allocating the fund value in the \GSAabbrevspace among the survivors, we can see that it is identical to the \PAFabbrevspace for a specific choice of the consumption rate in the latter.

Each member contributes 1 unit, so that the total fund value of the \GSAabbrevspace at the start of a year is $\numalive$ units.  Notionally allocating the total fund value among the members gives an initial notional fund value per member of
\[
F_{0}^{\textrm{\GSAabbrev}} = 1.
\]
At time 0, each member receives a payment of
\[
c_{0}^{\textrm{\GSAabbrev}} = \frac{F_{0}^{\textrm{\GSAabbrev}}}{\ddot{a}_{0}} = \frac{1}{\ddot{a}_{0}}.
\]
The remaining total \GSAabbrevspace fund value is $\numalive (1 - c_{0}^{\textrm{\GSAabbrev}})$ units.  As investment returns are assumed to be zero over the year, the total fund value in the \GSAabbrevspace at the end of the year remains $\numalive (1 - c_{0}^{\textrm{\GSAabbrev}})$ units.  Notionally allocating the end-of-year fund among the $\numalive - \numdead$ survivors gives a notional fund value per surviving member of
\begin{equation} \label{EQNgsanotfundvalue}
F_{1}^{\textrm{\GSAabbrev}} = \frac{\numalive (1 - c_{0}^{\textrm{\GSAabbrev}})}{\numalive - \numdead} =  \left( 1 + \frac{\numdead}{\numalive - \numdead} \right)( 1 - c_{0}^{\textrm{\GSAabbrev}} ).
\end{equation}
How does this compare with the notional fund per member in the \PAFabbrev? Comparing equations (\ref{EQNpafnotfundvalue}) and (\ref{EQNgsanotfundvalue}), we see that for a homogeneous group, the \GSAabbrevspace is identical to the \PAFabbrevspace with the specific choice of the consumption rate
\[
c_{0}^{\textrm{\PAFabbrev}} = c_{0}^{\textrm{\GSAabbrev}} = \frac{1}{\ddot{a}_{0}}.
\]

\subsection{Understanding the \GSAabbrevspace further}

Although the \GSAabbrevspace is a structure that allows individuals to pool their mortality risk together, it does not smooth investment returns over time.  We can illustrate the latter point by considering a group of individuals who are immortal.  Let the constant effective rate of interest used to calculate a life annuity payable to the immortals be $r > -1$, so that
\[
\ddot{a}_{n} = \sum_{k=0}^{\infty} \frac{1}{(1+r)^{k}} = \frac{1+r}{r}, \qquad \textrm{for $n=0,1,2,\ldots$}.
\]
Thus at time 0, in the \GSAabbrevspace the consumption per member at time 0 is
\[
c_{0}^{\textrm{\GSAabbrev}} = \frac{1}{\ddot{a}_{0}} = \frac{r}{1+r}.
\]
Suppose the \GSAabbrevspace fund achieves an investment return of $R_{1}>-1$ over the year through investment in a bond.  Then the total fund value grows to
\[
\numalive (1 - c_{0}^{\textrm{\GSAabbrev}}) (1 + R_{1}) = \numalive \frac{1 + R_{1}}{1+r},
\]
and notionally allocating it among the $\numalive$ immortals at time 1 gives an individual notional fund value of
\[
F_{1}^{\textrm{\GSAabbrev}} = \frac{1 + R_{1}}{1+r}.
\]
Hence the payment to each member at time 1 is
\[
c_{1}^{\textrm{\GSAabbrev}} = \frac{F_{1}^{\textrm{\GSAabbrev}}}{\ddot{a}_{1}} = \frac{r(1 + R_{1})}{(1+r)^{2}}.
\]
As $r$ is constant, the time 1 payment to each member could have been secured by investing the amount $\frac{r}{(1+r)^{2}}$ at time 0 in the bond.  Denoting by $R_{n}> -1$ the annual return achieved by the fund in year $n$ through investment in the bond, for $n=1,2,\ldots$, it follows that the consumption payment at time $n$ of
\[
\frac{r}{(1+r)^{n+1}} \prod_{k=1}^{n} (1 + R_{k})
\]
can be secured by investing the amount $\frac{r}{(1+r)^{n+1}}$ at time 0 in the bond.  This means that there is no investment smoothing in the \GSAabbrevspace across time.

\subsection{The \PAFabbrevspace is actuarially fair} \label{SUBSECactfairpaf}
We show that the \PAFabbrevspace is actuarially fair over each unit time period, which implies that it is actuarially fair over each individual's future lifetime. 
Without loss of generality, assume that there is no consumption at time 0 so that $c_{0}^{\textrm{\PAFabbrev}}=0$.   Specifically, we show that the expected value of an individual's benefits from the \PAFabbrevspace -- namely the fund value at time 1, $F_{1}^{\textrm{\PAFabbrev}}$ -- is equal to the individual's initial contribution $F_{0}^{\textrm{\PAFabbrev}} = 1$.  As there is nothing special about the first time period, the result generalizes to future time periods.  Note that we have set the interest rate to zero.

Denote by $\probsurvive \in (0,1)$ the probability that a member survives from time 0 to time 1, and set $\probdie = 1 - \probsurvive$.  We assume that if everyone in the fund dies at time 1 then their estates receive their individual fund values and there is no mortality risk-sharing.  As there are $\numalive$ members in the \PAFabbrevspace at time 0, such an event occurs with probability $\probdie^{\numalive}$.   With this assumption, the expected fund value at time 1 of a member is
\[
\begin{split}
\expectation \left( F_{1}^{\textrm{\PAFabbrev}} \right) & = \probdie^{\numalive} + \probsurvive \, \expectation \left( 1 + \frac{\numdead}{\numalive - \numdead} \, \bigg\vert \, \textrm{member survives to time 1} \right) \\
& = \probdie^{\numalive} + \probsurvive \sum_{n=0}^{\numalive-1} \frac{\numalive}{\numalive - n} \binom{\numalive-1}{n} \probdie^{n} \probsurvive^{\numalive - 1 - n} \\
& = 1 = F_{0}^{\textrm{\PAFabbrev}}.
\end{split}
\]
Thus the sharing rule for a \PAFabbrevspace is actuarially fair over the first unit time period, no matter how many members are in the group: two, ten, hundred or ten thousand.  It follows that the \PAFabbrevspace is actuarially fair over any unit time period, and hence it is actuarially fair over each individual's future lifetime.  Consequently, the homogeneous \GSAabbrevspace is also actuarially fair, since it is simply a \PAFabbrevspace with a specific choice of the consumption rate.

\section{Comparison of the \GSAabbrevspace to the \AOFabbrevspace for a homogeneous group}

After outlining the mortality risk-sharing mechanism in the \AOFspace (\AOFabbrev) for a homogeneous group of individuals, we compare the gains that a surviving member gets from the homogeneous \AOFabbrevspace to those in the homogeneous \GSAabbrev.  We show that, for a homogeneous group, the surviving members of the \AOFabbrevspace gain less from deaths than the survivors in the \GSAabbrev.

\subsection{The homogeneous \AOFabbrev}

The \AOFabbrevspace is presented in \citet{donnellyetal14.article}.  In its full theoretical generality, it allows individual members true consumption and investment freedom and they are free to leave the fund at any time. 

We continue to consider the homogeneous group described in Section \ref{SUBSEChomogroup}.  Each member has initial fund value $F_{0}^{\textrm{\AOFabbrev}} = 1$.  This is the actual fund value of each member, and not a notional fund value as in the \GSAabbrev.  While this may seem like an accounting matter, it is central to the different approaches of the \AOFabbrevspace and the \GSAabbrev.  In the \AOFabbrev, members have absolute control over their own fund values.  In the \GSAabbrev, members make an initial contribution in exchange for a future consumption stream.  They do not own a notional fund value.

To allow a sensible comparison with the \GSAabbrev, suppose that members of the \AOFabbrevspace consume the same amount $c_{0}^{\textrm{\AOFabbrev}}$ of their fund value at time 0, leaving each with wealth
\[
F_{0+}^{\textrm{\AOFabbrev}} = 1 - c_{0}^{\textrm{\AOFabbrev}}.
\]
Upon each death, the wealth of the dead individual is shared among the group members who were alive just before the time of death.  The share that each member receives is proportional to their individual mortality rate and individual fund value.  However, as members are identical in the homogeneous group, this simplifies to each member receiving an equal share.  For example, suppose there are 10 people in the homogeneous group, each with wealth 500 units at the start of a day.  One of the group dies during the day.  Then at the end of the day, each of the 10 members receives 50 units due to the death.  Thus each of the 9 survivors has wealth 550 units at the end of the day, and the estate of the dead member receives a payment of 50 units.

Suppose the first death in the group occurs just before time $T_{1}$.  In the \AOFabbrev, the wealth of the dead member is shared among everyone who was alive at time $T_{1-}$, and not only among the survivors at the time of the individual's death.  As it is a homogeneous group and investment returns are assumed to be zero, each person who was alive at time $T_{1-}$ receives the amount
\[
\frac{1}{\numalive} F_{T_{1}-}^{\textrm{\AOFabbrev}}
\]
at time $T_{1}$, giving each surviving member a total fund value of
\[
F_{T_{1}}^{\textrm{\AOFabbrev}} = F_{T_{1}-}^{\textrm{\AOFabbrev}} + \frac{1}{\numalive} F_{T_{1}-}^{\textrm{\AOFabbrev}} = \left( 1 + \frac{1}{\numalive} \right) (1 - c_{0}^{\textrm{\AOFabbrev}} )
\]
at time $T_{1}$.  The first dead member has a fund value at time $T_{1}$ of $F_{T_{1}-}^{\textrm{\AOFabbrev}} / \numalive$, which is paid to their estate.  This member is then no longer part of the \AOFabbrevspace so that the group consists of $\numalive - 1$ members at time $T_{1}$.
 
Suppose the second death in the group occurs just before time $T_{2} > T_{1}$.  The wealth of this dead member is shared among the $\numalive - 1$ members who were alive at time $T_{2-}$ (and not among those who were alive at time 0).  Thus each member who was alive at time $T_{2-}$ receives the amount
\[
\frac{1}{\numalive - 1} F_{T_{2}-}^{\textrm{\AOFabbrev}},
\]
resulting in each surviving member having a total fund value of
\[
F_{T_{2}}^{\textrm{\AOFabbrev}} = F_{T_{2}-}^{\textrm{\AOFabbrev}} + \frac{1}{\numalive - 1} F_{T_{2}-}^{\textrm{\AOFabbrev}} = \left( 1 + \frac{2}{\numalive - 1} \right)  (1 - c_{0}^{\textrm{\AOFabbrev}} ).
\]
The second dead member has a fund value at time $T_{2}$ equal to $F_{T_{2}-}^{\textrm{\AOFabbrev}} / (\numalive - 1)$, which is paid to their estate.  This member is no longer part of the group so that the group consists of $\numalive - 2$ members at time $T_{2}$.

In summary, we find that if there are a total of $\numdead$ deaths over the first time period in a homogeneous \AOFabbrev, then each surviving member at time 1 has a total fund value of
\[
F_{1}^{\textrm{\AOFabbrev}} = \left( 1 + \frac{\numdead}{\numalive + 1 - \numdead} \right) (1 - c_{0}^{\textrm{\AOFabbrev}} ).
\]

\subsection{Discussion of the \GSAabbrevspace compared to the \AOFabbrevspace for a homogeneous group}

The \GSAabbrevspace and the \AOFabbrevspace are both structures that allow individuals to pool their mortality risk together.  Neither of them allows investment risk to be shared across time; members bear their own investment risk.

However, their operation is fundamentally different.  In the \GSAabbrev, the calculation and timing of the consumption payments to the members are fixed in advance.  The members do not have individual investment freedom, but must as a group decide how to invest their funds.  Moreover, the intention of the \GSAabbrevspace is that, once members have joined, they cannot exit before death.

In comparison, the members of the \AOFabbrevspace can withdraw as much as they like of their own funds, whenever they like.  Although in practice reducing the potential impact of moral hazard and adverse selection are likely to require some restrictions on the amount and timing of withdrawals (this is discussed further in \citet{donnellyetal14.article}), the \emph{modus operandi} of the \AOFabbrevspace is to be actuarially fair over every instant in time for every member, rather than over a member's lifetime only or over an entire group.  The implication of maintaining actuarial fairness over all instants in time is that members can withdraw all of their funds from the \AOFabbrev, and thus exit the \AOFabbrevspace before their deaths.  It means that they can invest their money however they choose: one member can invest in bonds, the other in property and a third in equities.

However, for a homogeneous group the greater flexibility of the \AOFabbrevspace over the \GSAabbrevspace has a financial consequence for the surviving members of the fund.  A surviving member at time 1 in the homogeneous \AOFabbrevspace who consumes the same amount at time 0 as in the homogeneous \GSAabbrev, will have a fund value at time 1 that is smaller than her notional fund value in the \GSAabbrev:
\[
F_{1}^{\textrm{\AOFabbrev}} - F_{1}^{\textrm{\GSAabbrev}} = - \frac{\numdead}{(\numalive - \numdead)(\numalive + 1 - \numdead)} (1 - c_{0}^{\textrm{\AOFabbrev}} ) \qquad \Rightarrow \qquad F_{1}^{\textrm{\AOFabbrev}} \leq F_{1}^{\textrm{\GSAabbrev}}, \qquad \textrm{a.s}.
\]
As the number of members $\numalive$ increases, the difference $F_{1}^{\textrm{\AOFabbrev}} - F_{1}^{\textrm{\GSAabbrev}}$ diminishes to zero.  Indeed, as shown in \citet{donnellyetal13.article}, as the number of members tends to infinity, the \AOFabbrevspace tends to the \PAFabbrev.  Thus for an infinite number of homogeneous members, the two structures, the \AOFabbrevspace and the \GSAabbrev, coincide.

Interestingly, the inequality $F_{1}^{\textrm{\AOFabbrev}} \leq F_{1}^{\textrm{\GSAabbrev}}$ does not continue to hold true a.s. for every survivor in a heterogeneous group.  We show this in Section \ref{SUBSECgsaaofhetero}.

\section{The heterogeneous \GSAabbrev}

Here we show that the heterogeneous \GSAabbrevspace is not actuarially fair over a member's lifetime.  For example, the expected lifetime benefits received by wealthy members are less than their initial contributions.  We begin by calculating the consumption payments made by \GSAabbrevspace in a simple two-period model and then discuss the lack of lifetime actuarial fairness.

\subsection{The heterogeneous group} \label{SUBSEChetergroup}

Suppose that there are two distinct groups of people: Group $A$ and Group $B$.  Group $A$ consists of $\numalive^{A} > 1$ members, each of whom contributes $\notfundA_{0} > 0$ to the fund.  Group $B$ contains $\numalive^{B} > 1$ members, each of whom contributes $\notfundB_{0} > 0$ to the fund. 

We use a simple two-period model that is enough to illustrate the results.  Suppose that each individual in Group $X \in \{A,B\}$ who is alive at time 0 has probability $\probsurviveX \in (0,1)$ of surviving to time 1.  All individuals are dead by time 2 and deaths occur independently of each other.

For simplicity, we continue to assume that the interest rate is zero.  Thus the expected present value of a life annuity payment of 1 unit per time period is
\[
\ddot{a}_{0}^{\textrm{X}} := 1 + \probsurviveX \textrm{ at time 0, and } \ddot{a}_{1}^{\textrm{X}} := 1 \textrm{ at time 1}, \qquad \textrm{for $X \in \{A,B\}$}.
\]

\subsection{The heterogeneous \GSAabbrev} \label{SUBSECheterogsa}
Here we calculate the consumption payments made by \GSAabbrevspace to the members of the above heterogeneous group.  We include the calculations for our specific case for the sake of completeness: the more general calculations, which allow for more subgroups and non-zero interest rates, are found in \citet{piggottetal05.article}.

Every member in Group $X$ receives a consumption payment at time 0 of 
\begin{equation} \label{EQNgsaconsum}
\consumXzero := \frac{\notfundX_{0}}{\ddot{a}_{0}^{X}},
\end{equation}
leaving each member with notional fund value $\notfundX_{0+} := \notfundX_{0} - \consumXzero$ for each $X \in \{ A, B \}$.

At time 1, the total fund value in the \GSAabbrevspace of $\notfundA_{0+} \numalive^{A} + \notfundB_{0+} \numalive^{B}$ is allocated notionally among the survivors according to a pre-determined rule.  Let the random variable $\numdead^{A}$ denote the number of dead in Group $A$ by time 1, and let the random variable $\numdead^{B}$ denote the number of dead in Group $B$ by time 1.  Thus there are $\numalive^{A} - \numdead^{A}$ survivors in Group $A$ and $\numalive^{B} - \numdead^{B}$ survivors in Group $B$ at time 1.

Denoting by $\fundfraction$ the mortality experience adjustment factor of \citet{piggottetal05.article}, with
\begin{equation} \label{EQNmea}
\fundfraction := \frac{\notfundA_{0+} \numalive^{A} + \notfundB_{0+} \numalive^{B}}{\notfundA_{0+} \frac{\numalive^{A} - \numdead^{A}}{\probsurviveA} + \notfundB_{0+} \frac{\numalive^{B} - \numdead^{B}}{\probsurviveB}},
\end{equation}
then for each $X \in \{ A,B \}$, the notional fund value of each survivor in Group $X$ jumps to
\[
\notfundX_{1} = \fundfraction \cdot \consumXzero.
\]
As death is certain to occur by time 2, then a survivor consumes all of their notional fund value at time 1, i.e.
\begin{equation} \label{EQNhetgsaconsumX}
\consumXone := \notfundX_{1} = \fundfraction \cdot \consumXzero.
\end{equation}

Equations (\ref{EQNmea}) and (\ref{EQNhetgsaconsumX}) demonstrate the fundamental idea of the mortality risk-sharing rule underlying the heterogeneous \GSAabbrev: the consumption payments of all individuals in the \GSAabbrevspace change by the same ratio according to how far the number of survivors in each group deviate from the expected number, weighted by the fund value of each group member.  If the actual number of deaths in each group turns out as expected then there is no change in the consumption payments (ignoring investment returns).

Notice that the calculation of the initial consumption payment (equation (\ref{EQNgsaconsum})) does not involve the mortality experience adjustment factor (equation (\ref{EQNmea})).  This is the reason why the \GSAabbrevspace is not actuarially fair when there is a finite number of members.  It is only when mortality risk is fully diversified, i.e. with infinitely-many people in each group, that $\fundfraction = 1$ almost surely and the initial consumption payment $\consumXzero$ equals the expected consumption payment.


When mortality risk is not fully diversified, i.e. with only finitely-many people in the \GSAabbrev, the heterogeneous \GSAabbrevspace is not actuarially fair over an an individual's lifetime in the sense that
\[
\notfundX_{0} \neq \mathbb{E} \left( \consumXzero + \consumXone \right),
\]
for each $X \in \{A,B\}$.  We show this next.

\subsection{The heterogeneous \GSAabbrevspace is not actuarially fair} \label{SUBSECactfairheteromort}

As before, we define a fund as actuarially fair if the expected value at time 0 of an individual's benefits from the fund is equal to their contributions.   We show that for a heterogeneous group with a finite number of members, the \GSAabbrevspace of \citet{piggottetal05.article} is not actuarially fair.  The poorer members of the scheme benefit from subsidies from the richer members, and the young benefit from subsidies from the old.

Fix a member in Group $B$, whom we call \namembrB.  We assume that if everyone in the \GSAabbrevspace dies at time 1, their estates receive their individual time 0 notional fund values.  Such an event occurs with probability $(\probdieA)^{\numalive^{A}} (\probdieB)^{\numalive^{B}}$.

If the \GSAabbrevspace is actuarially fair, then \namembrBspace should have expected consumption equal to his initial contribution of $\notfundB_{0}$, i.e.
\begin{equation} \label{EQNhetgsaactfairB}
\expectation (\consumBzero + \consumBone ) = \notfundB_{0},
\end{equation}
regardless of the wealth-mortality characteristics of Group $A$ and Group $B$, each represented by the triple
\[
\grpXchars = (\numalive^{X}, \probsurviveX, \notfundX_{0})
\]
in which $\numalive^{X}$ is the number of members in Group $X$ at time 0, $\probsurviveX$ is the probability of an individual in Group $X$ surviving to time 1 and $\notfundX_{0}$ is the wealth at time 0 of a member of Group $X$, for each $X \in \{A,B\}$.  We have already seen in Section \ref{SUBSECheterogsa} that equation (\ref{EQNhetgsaactfairB}) holds for a \GSAabbrevspace in which $\grpAchars = (\infty, \probsurviveA, \notfundA_{0})$ and $\grpBchars = (\infty, \probsurviveB, \notfundB_{0})$.

However, as we show in Proposition \ref{PROPNsamesizesgsa}, with the group wealth-mortality characteristics
\[
\grpAchars = (\numalive, \probsurvive, 1) \quad \textrm{and} \quad \grpBchars = (\numalive, \probsurvive, \notfundB_{0}),
\]
for constants $\numalive \in \naturalnumbers$ and $\probsurvive \in (0,1)$, we find that 
\[
\expectation (\consumBzero + \consumBone) < \notfundB_{0}, \qquad \textrm{for $\notfundB_{0} > 1$}.
\]
This means that \namembrB, who is wealthier than each member of Group $A$, has an expected consumption in the \GSAabbrevspace that is less than his initial contribution of $\notfundB_{0}$.  \namembrBspace can obtain a higher expected consumption, equal to his initial contribution of $\notfundB_{0}$, by joining a homogeneous \GSAabbrevspace in which everyone is an independent copy of him.

We also show in Proposition \ref{PROPNsamesizesgsa} that an analogous result continues to hold if we condition upon \namembrB 's survival to time 1: \namembrBspace obtains a higher consumption from a homogeneous \GSAabbrevspace than from the heterogeneous \GSAabbrev.  Thus if \namembrBspace does not care what happens when he dies, then he still has a higher expected consumption from the homogeneous \GSAabbrev, in which everyone is an independent copy of him, than from the heterogeneous \GSAabbrev, in which members of Group $A$ contribute less than members of Group $B$.

\begin{restatable}[]{proposition}{samesize} \label{PROPNsamesizesgsa}
Consider a heterogeneous \GSAabbrevspace consisting only of two groups, $A$ and $B$, with the same number of members in each group, i.e. $\numalive := \numalive^{A} = \numalive^{B}$.  Each member of Group $A$ has initial wealth $\notfundA_{0}=1$ and each member of Group $B$ has initial wealth $\notfundB_{0} > 0$.

Fix a member in Group $B$, whom we call \namembrB.  For $\notfundB_{0} > 1$,
\begin{itemize}
	\item \namembrB 's expected consumption in the heterogeneous \GSAabbrevspace is less than his initial wealth $\notfundB_{0}$, and
	\item \namembrB 's expected consumption conditional upon his survival to time 1 in the heterogeneous \GSAabbrevspace is less than in a homogeneous \GSAabbrev, in which all surviving members have the same amount of wealth as \namembrB.
\end{itemize}
For $\notfundB_{0}  < 1$,
\begin{itemize}
	\item \namembrB 's expected consumption in the heterogeneous \GSAabbrevspace is greater than his initial wealth $\notfundB_{0}$, and
	\item \namembrB 's expected consumption conditional upon his survival to time 1 in the heterogeneous \GSAabbrevspace is greater than in a homogeneous \GSAabbrev, in which all surviving members have the same amount of wealth as \namembrB.
\end{itemize}
\end{restatable}

\begin{proof}
See the Appendix.
\end{proof}

We show in Proposition \ref{PROPNbobonlysgsa} that the actuarial unfairness continues to hold in a \GSAabbrevspace in which \namembrBspace is the only member of Group $B$ and there is more than one member of Group $A$, i.e. with group wealth-mortality characteristics
\[
\grpAchars = (\numalive, \probsurvive, 1) \quad \textrm{and} \quad \grpBchars = (1, \probsurvive, \notfundB_{0}),
\]
for constants $\numalive \in \{2,3,\ldots,\}$ and $\probsurvive \in (0,1)$.

\begin{restatable}[]{proposition}{bobonly} \label{PROPNbobonlysgsa}
Consider a heterogeneous \GSAabbrevspace consisting only of two groups, $A$ and $B$.  Suppose there are $\numalive^{A} > 1$ members in Group $A$ and only one member, called \namembrB, in Group $B$.  Each member of Group $A$ has initial wealth $\notfundA_{0}=1$ and \namembrBspace has initial wealth $\notfundB_{0} > 0$.

For $\notfundB_{0} > 1$,
\begin{itemize}
	\item \namembrB 's expected consumption in the heterogeneous \GSAabbrevspace is less than his initial wealth $\notfundB_{0}$, and
	\item \namembrB 's expected consumption conditional upon his survival to time 1 in the heterogeneous \GSAabbrevspace is less than in a homogeneous \GSAabbrev, in which all surviving members have the same amount of wealth as \namembrB.
\end{itemize}
For $\notfundB_{0} < 1$,
\begin{itemize}
	\item \namembrB 's expected consumption in the heterogeneous \GSAabbrevspace is greater than his initial wealth $\notfundB_{0}$, and
	\item \namembrB 's expected consumption conditional upon his survival to time 1 in the heterogeneous \GSAabbrevspace is greater than in a homogeneous \GSAabbrev, in which all surviving members have the same amount of wealth as \namembrB.
\end{itemize}
\end{restatable}

\begin{proof}
See the Appendix.
\end{proof}

Through numerical evaluation, we demonstrate that the \GSAabbrevspace is actuarially unfair when the members of Group $A$ have different survival probabilities to the members of Group $B$.  We find that the older (or unhealthy) members have an expected consumption that is lower than their initial contribution.  The younger (or healthy) members have an expected consumption that is higher than their initial contribution. 

\begin{figure}[ht]
 \centering
 \subfigure[10 members in each group.]{
  \includegraphics[height=60mm,width=60mm, angle=270]{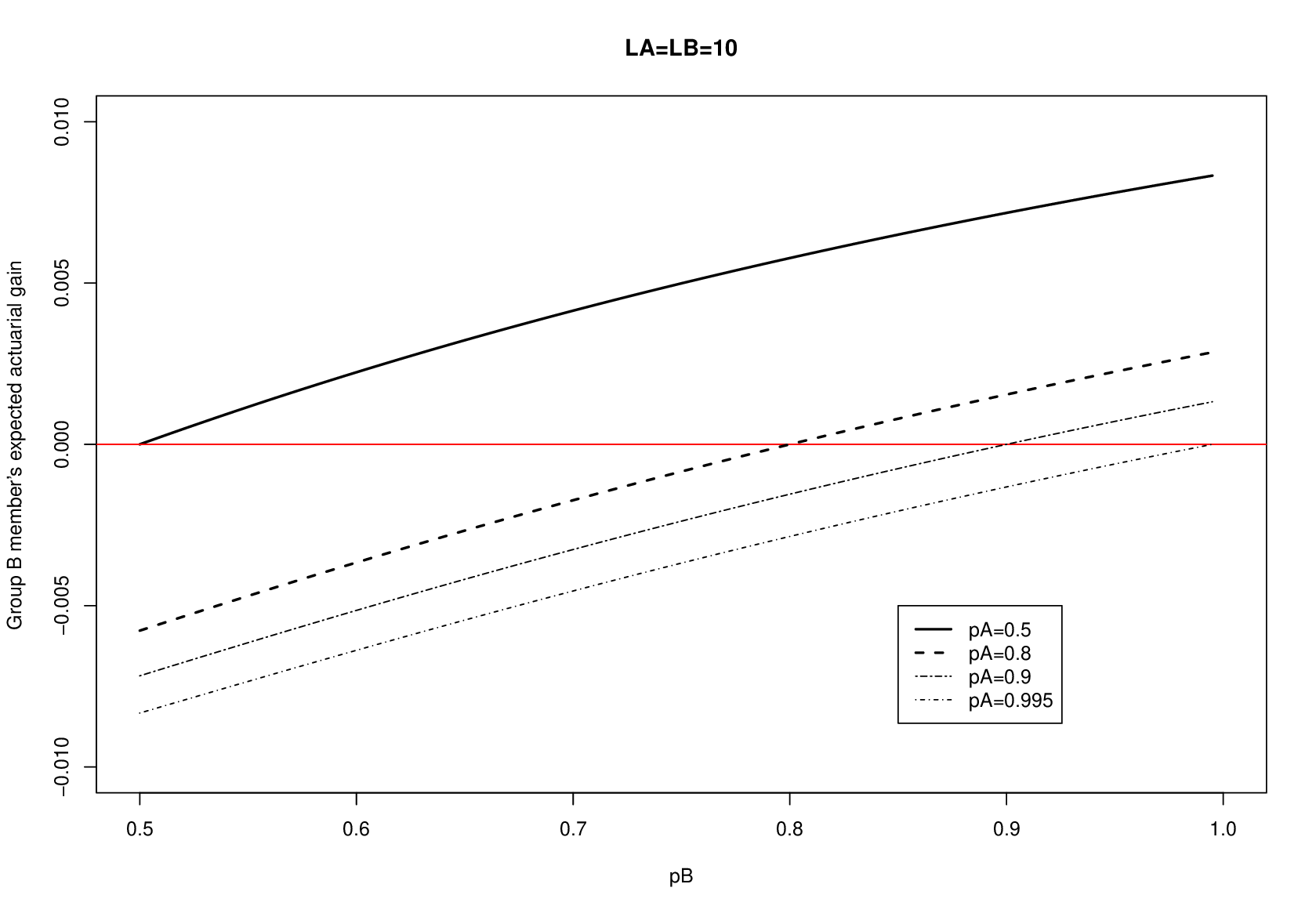}
   }
 \subfigure[100 members in each group.]{
  \includegraphics[height=60mm,width=60mm, angle=270]{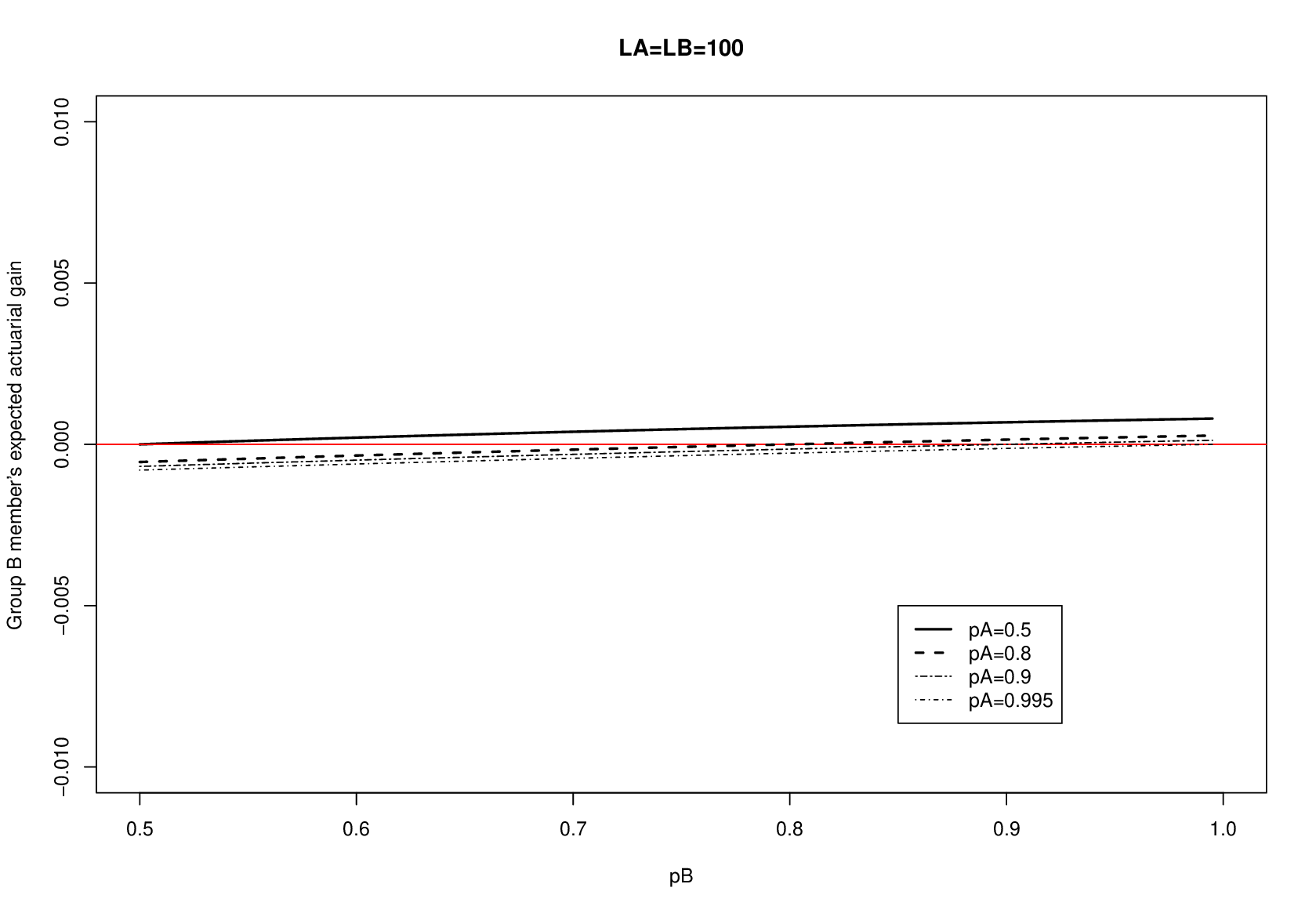}
   }
 \subfigure[10 members in Group A, 1 member in Group B.]{
  \includegraphics[height=60mm,width=60mm, angle=270]{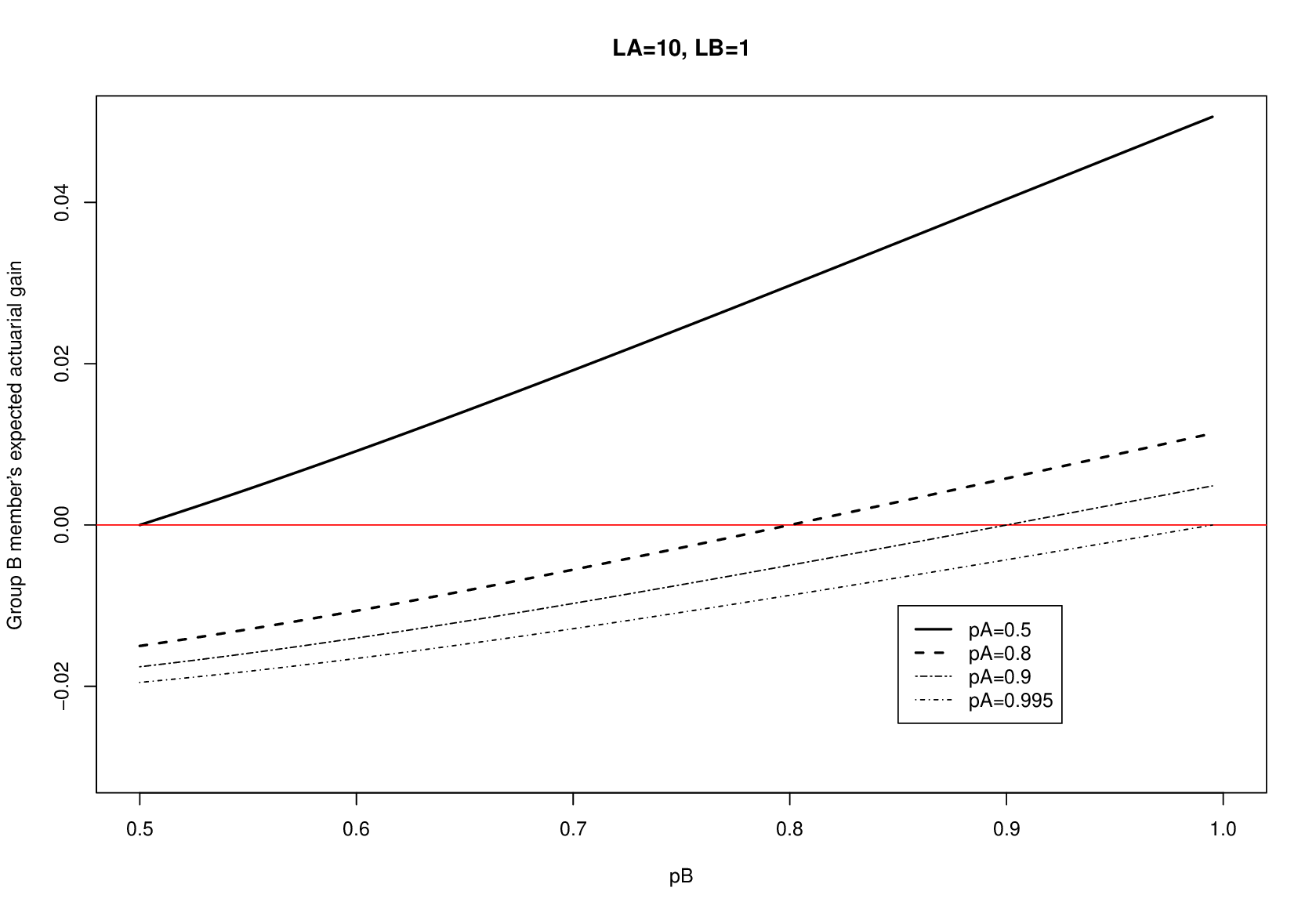}
   }
	 \subfigure[100 members in Group A, 1 member in Group B.]{
  \includegraphics[height=60mm,width=60mm, angle=270]{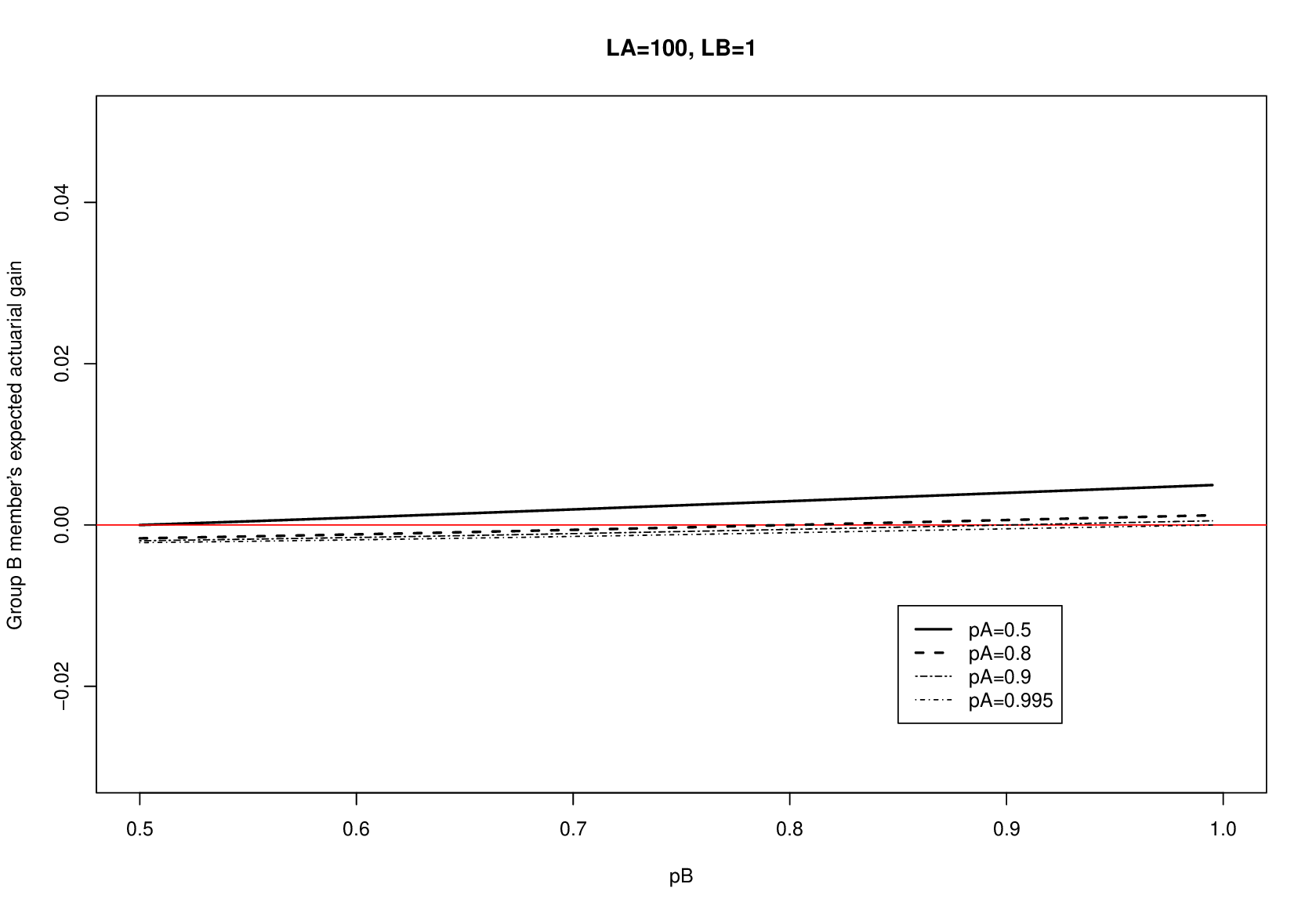}
   }
 \caption{The expected actuarial gain per unit initial contribution for a Group $B$ member as the survival probability $\probsurviveB$ varies in a heterogeneous \GSAabbrev.  The figures show the gain for a fixed number of members in each group, fixed survival probabilities $\probsurviveA$ and the same initial contribution of 1 unit from each Group $A$ and Group $B$ member.}
 \label{FIGactgainssurv}
\end{figure}

For a \GSAabbrevspace with different group survival probabilities, we assume that all members contribute initially 1 unit to the fund and calculate
\[
\expectation (\consumBzero + \consumBone) - 1,
\]
which is the expected actuarial gain from sharing mortality risk.  In an actuarially fair fund, the net expected actuarial gain is zero for all members.  If it is greater than zero for a member, then the member has an expected consumption that is higher than their initial contribution of 1 unit, and vice versa.  The results, illustrated in Figure \ref{FIGactgainssurv}, show that
\[
\expectation (\consumBzero + \consumBone) - 1 \left\{ \begin{array}{ll}
< 0  & \textrm{if $\probsurviveA > \probsurviveB$} \\
> 0 & \textrm{if $\probsurviveA < \probsurviveB$},
\end{array} \right.
\]
i.e. the \GSAabbrevspace with different group survival probabilities is not actuarially fair.

\subsection{Is actuarial unfairness financially significant?}

We have shown that the \GSAabbrevspace is actuarially unfair.  Here we investigate if the fairness is financially significant.  We find that it can be for small or highly heterogeneous groups.  \citet{qiaosherris12.article} analyze numerically the number of members required to adequately pool mortality risk in a heterogeneous \GSAabbrev.

First we calculate \namembrB 's expected actuarial gain per unit initial contribution in a heterogeneous \GSAabbrev, i.e.
\[
\frac{\expectation (\consumBzero + \consumBone) - \notfundB_{0}}{\notfundB_{0}}
\]
(in an actuarially-fair fund, the expected actuarial gain is zero for each member).

\begin{figure}[ht]
 \centering
 \subfigure[10 members in each group.]{
  \includegraphics[height=60mm,width=60mm, angle=270]{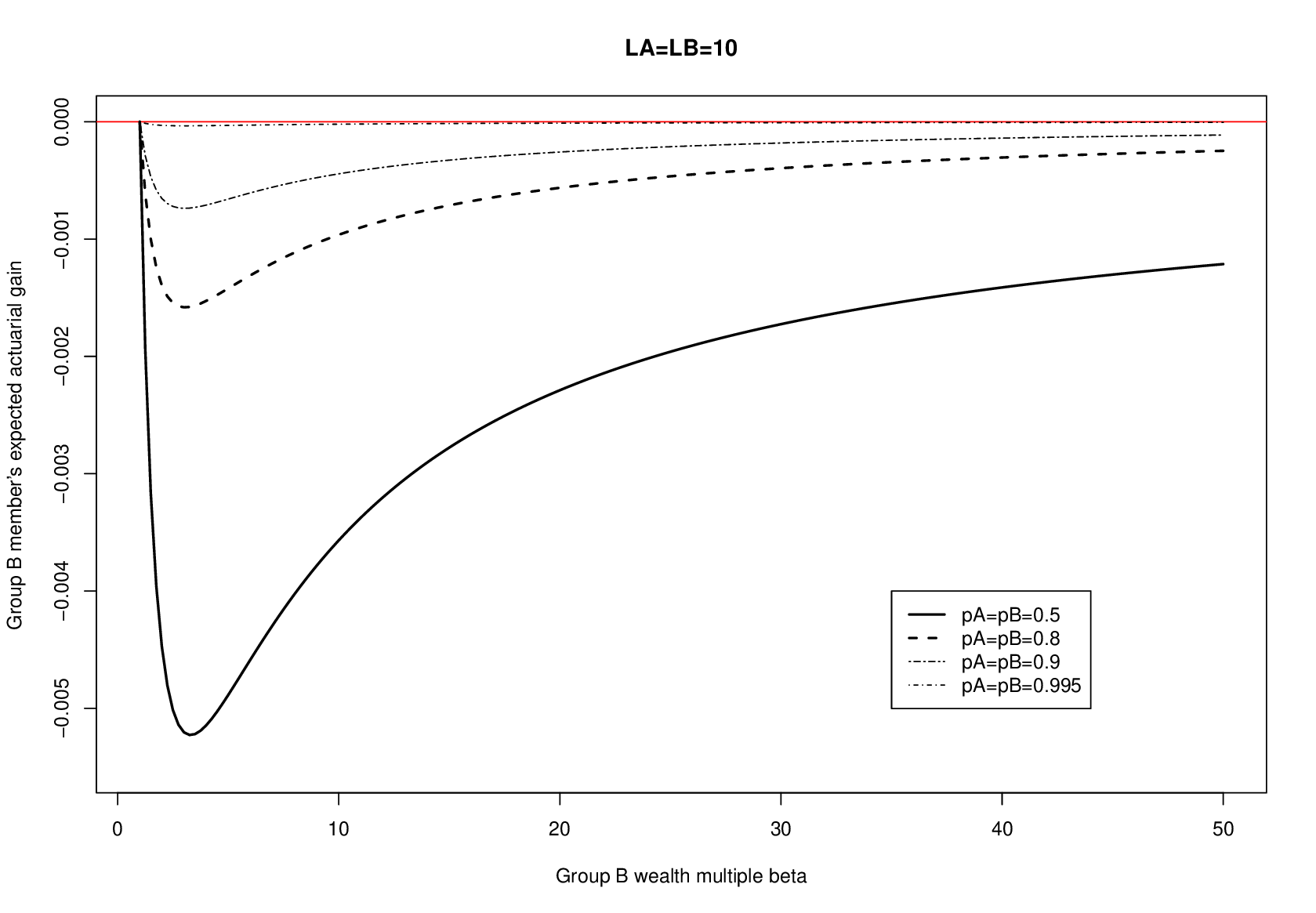}
   }
 \subfigure[100 members in each group.]{
  \includegraphics[height=60mm,width=60mm, angle=270]{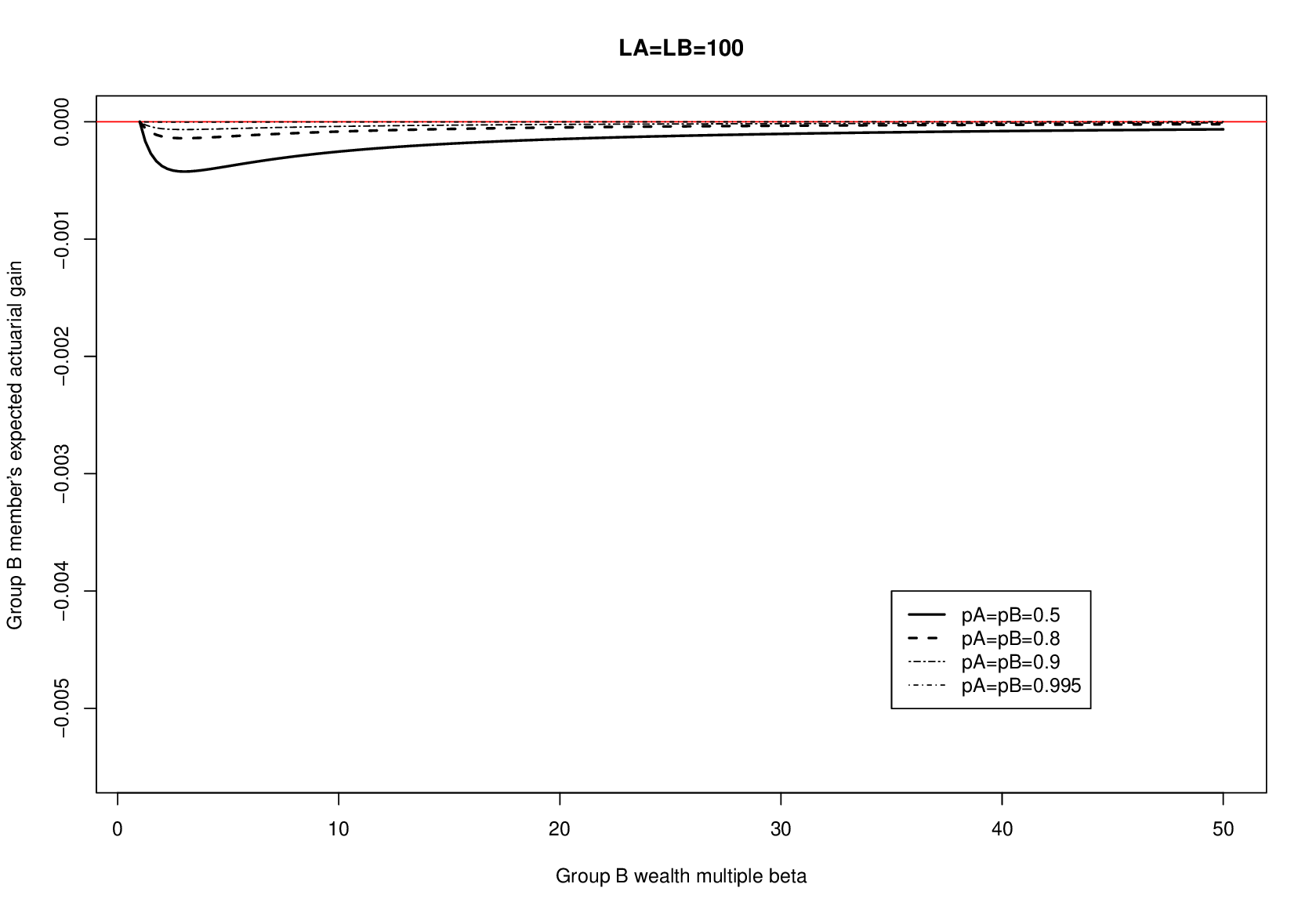}
   }
 \subfigure[10 members in Group A, 1 member in Group B.]{
  \includegraphics[height=60mm,width=60mm, angle=270]{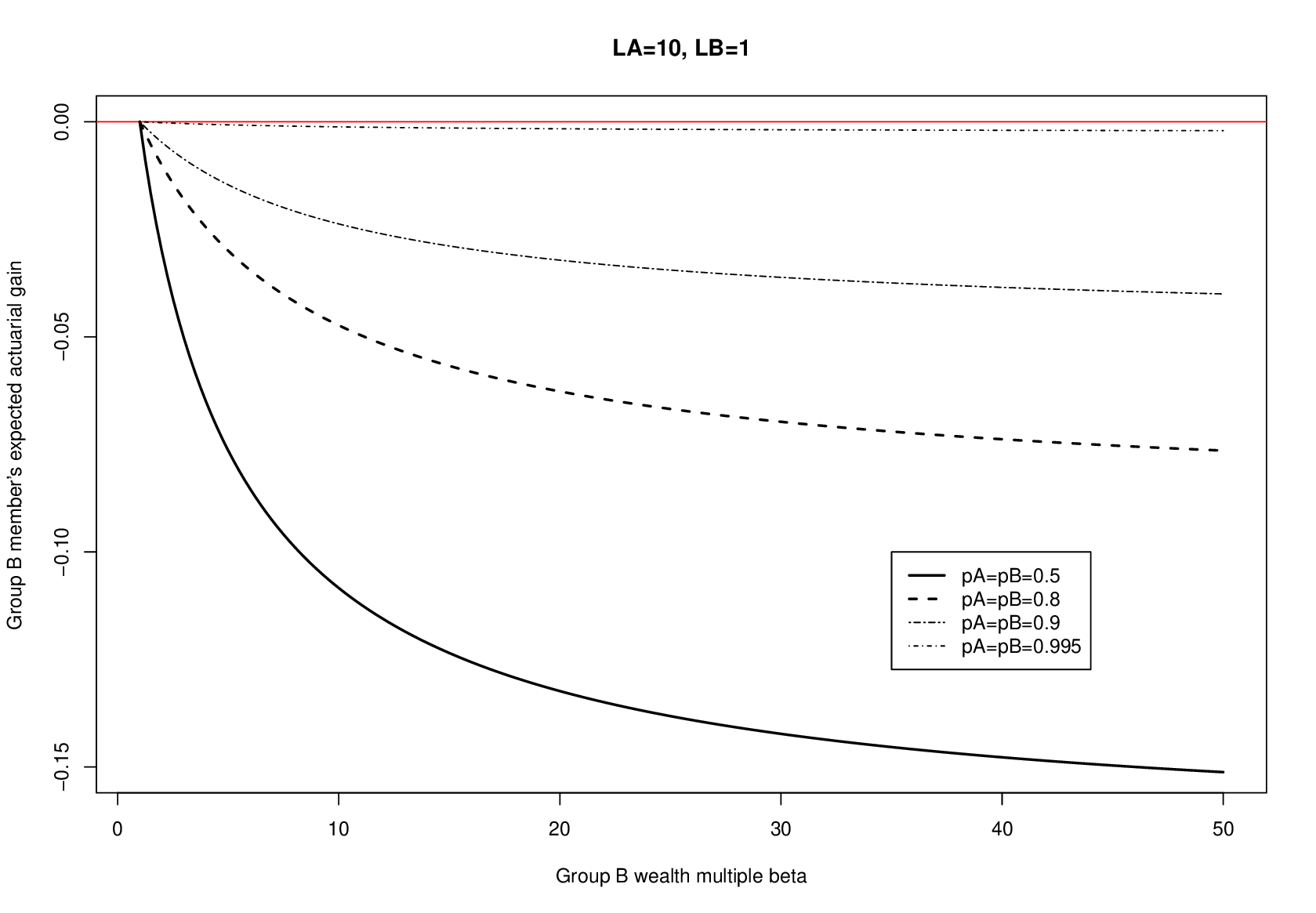}
   }
	 \subfigure[100 members in Group A, 1 member in Group B.]{
  \includegraphics[height=60mm,width=60mm, angle=270]{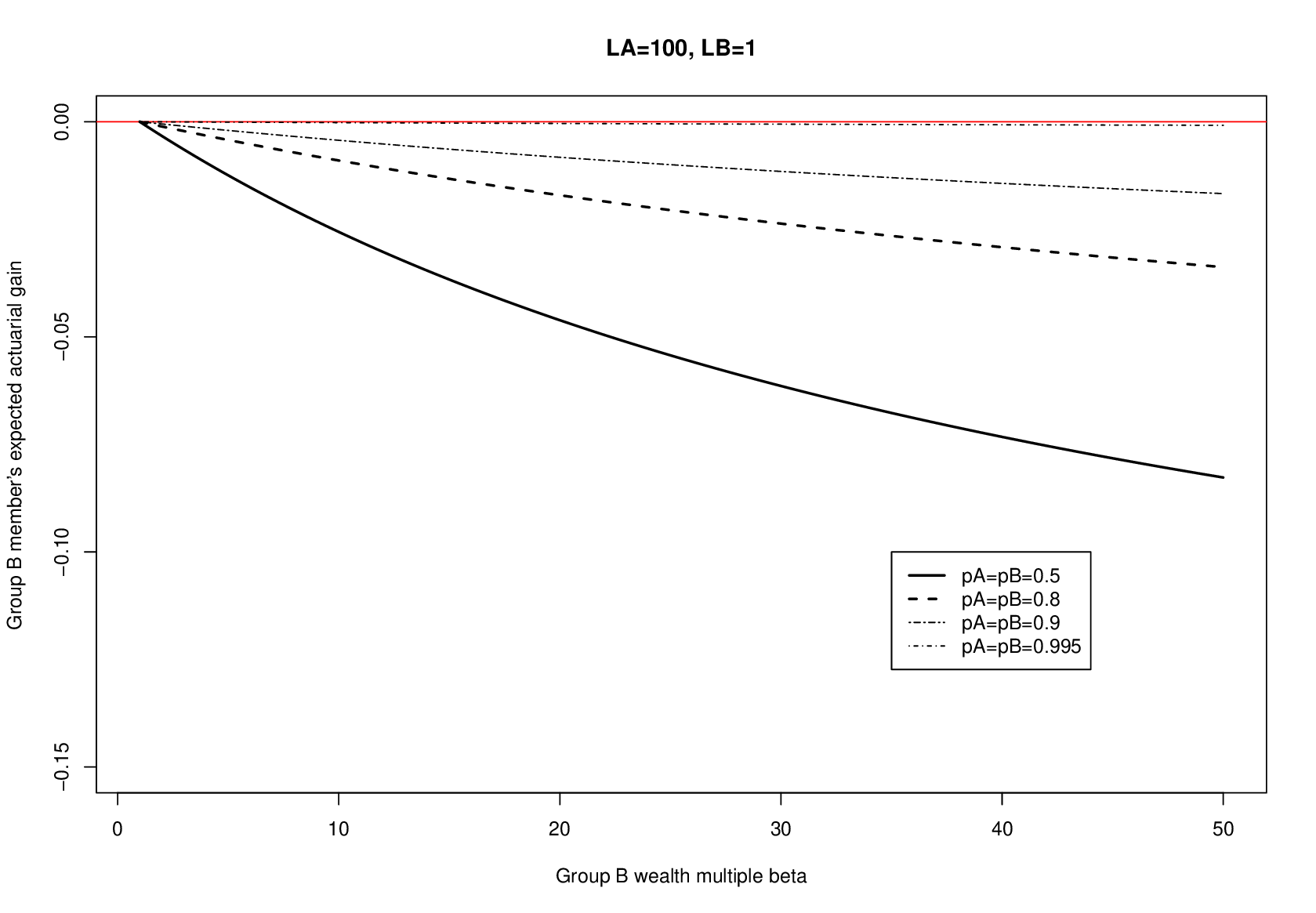}
   }
 \caption{The expected actuarial gain per unit initial contribution for a Group B member as the initial contribution of the Group $B$ member varies, in a heterogeneous \GSAabbrev.  The figures show the gain for a fixed number of members in each group, fixed survival probabilities $\probsurviveA$ and $\probsurviveB$, and initial contribution of 1 unit from each Group $A$ member.}
 \label{FIGactgainscont}
\end{figure}

The expected actuarial gain is shown in Figures \ref{FIGactgainssurv} and \ref{FIGactgainscont} for various sample heterogeneous \GSAabbrev s.  The results show that it is only with very small or highly heterogeneous groups that \namembrB 's expected actuarial loss is significant.   For example, for $\probsurviveA=\probsurviveB=0.9$, \namembrBspace the only member of Group $B$ and \namembrBspace making 10 times the initial contribution of each Group $A$ member, \namembrB 's expected actuarial gain is $-2.4\%$ with 10 members in Group $A$, but this declines to $-0.4\%$ with 100 members in Group $A$.

Next we turn to \namembrB 's gain conditional upon his survival to time 1, relative to a homogeneous \GSAabbrevspace in which everyone has the same probability of survival as \namembrB.  If the conditional gains are greater than zero, then \namembrBspace can obtain a higher consumption while he is alive from the heterogeneous \GSAabbrev, and vice versa.

\begin{figure}[ht]
 \centering
 \subfigure[10 members in each group.]{
  \includegraphics[height=60mm,width=60mm, angle=270]{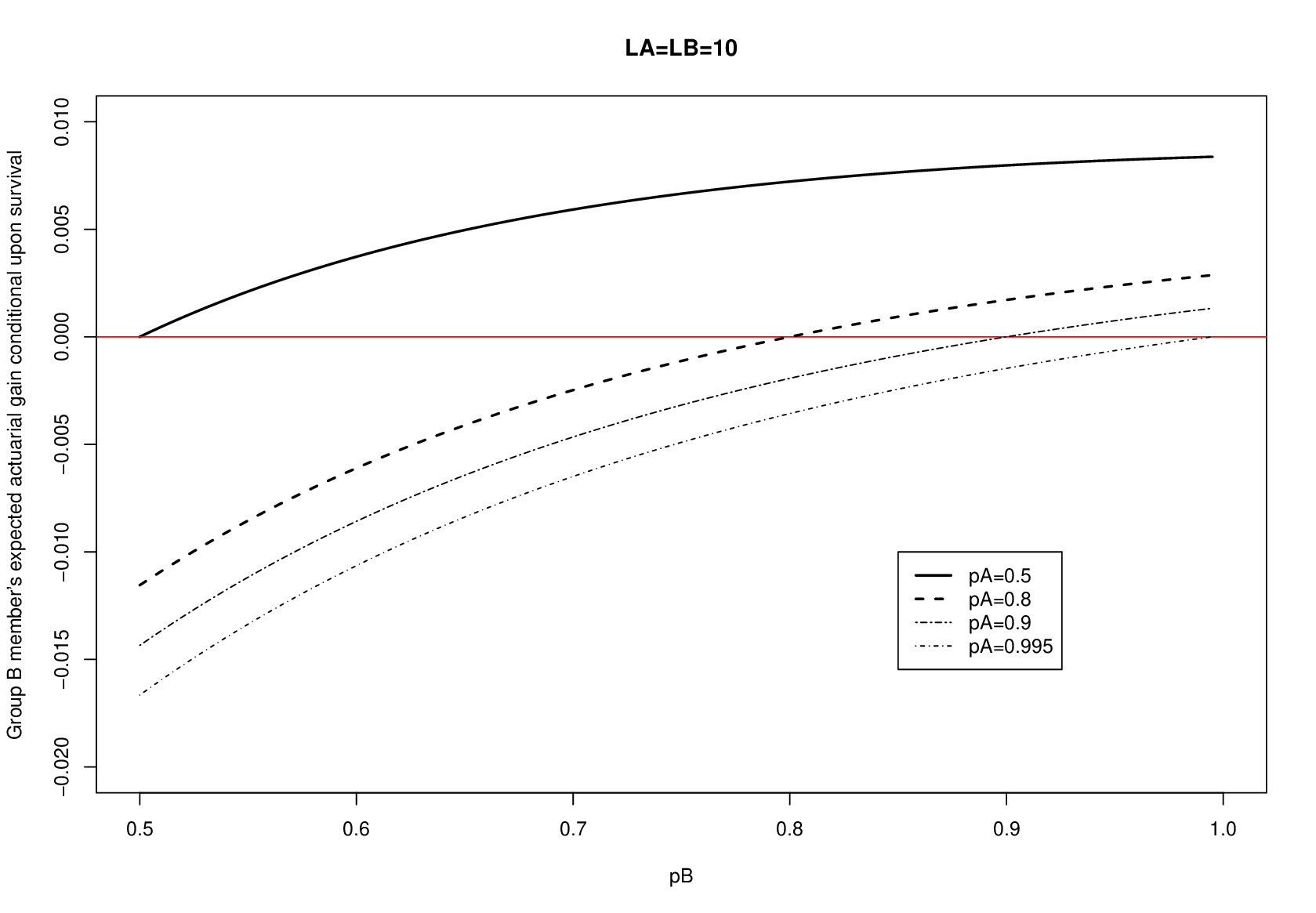}
   }
 \subfigure[100 members in each group.]{
  \includegraphics[height=60mm,width=60mm, angle=270]{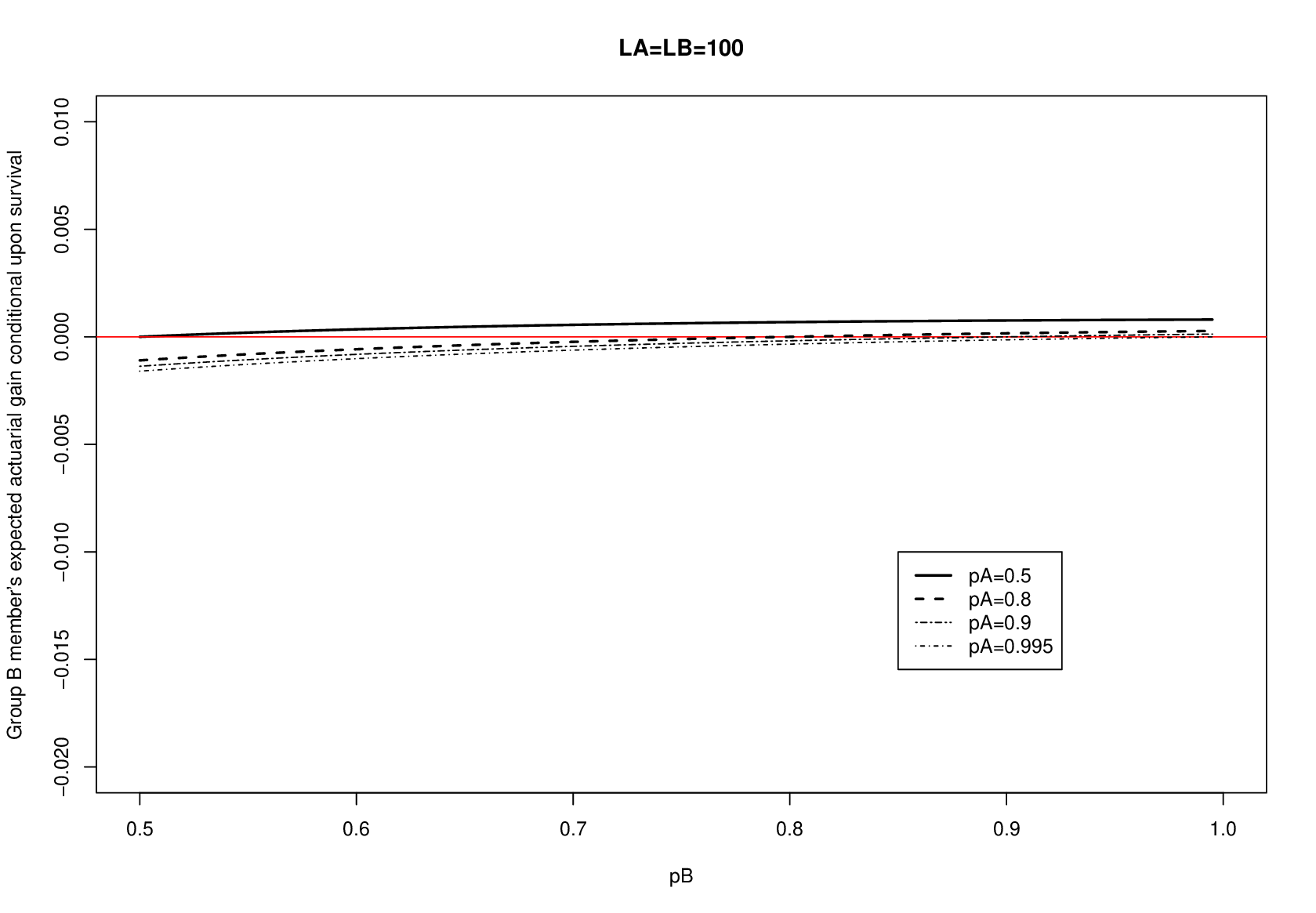}
   }
 \subfigure[10 members in Group A, 1 member in Group B.]{
  \includegraphics[height=60mm,width=60mm, angle=270]{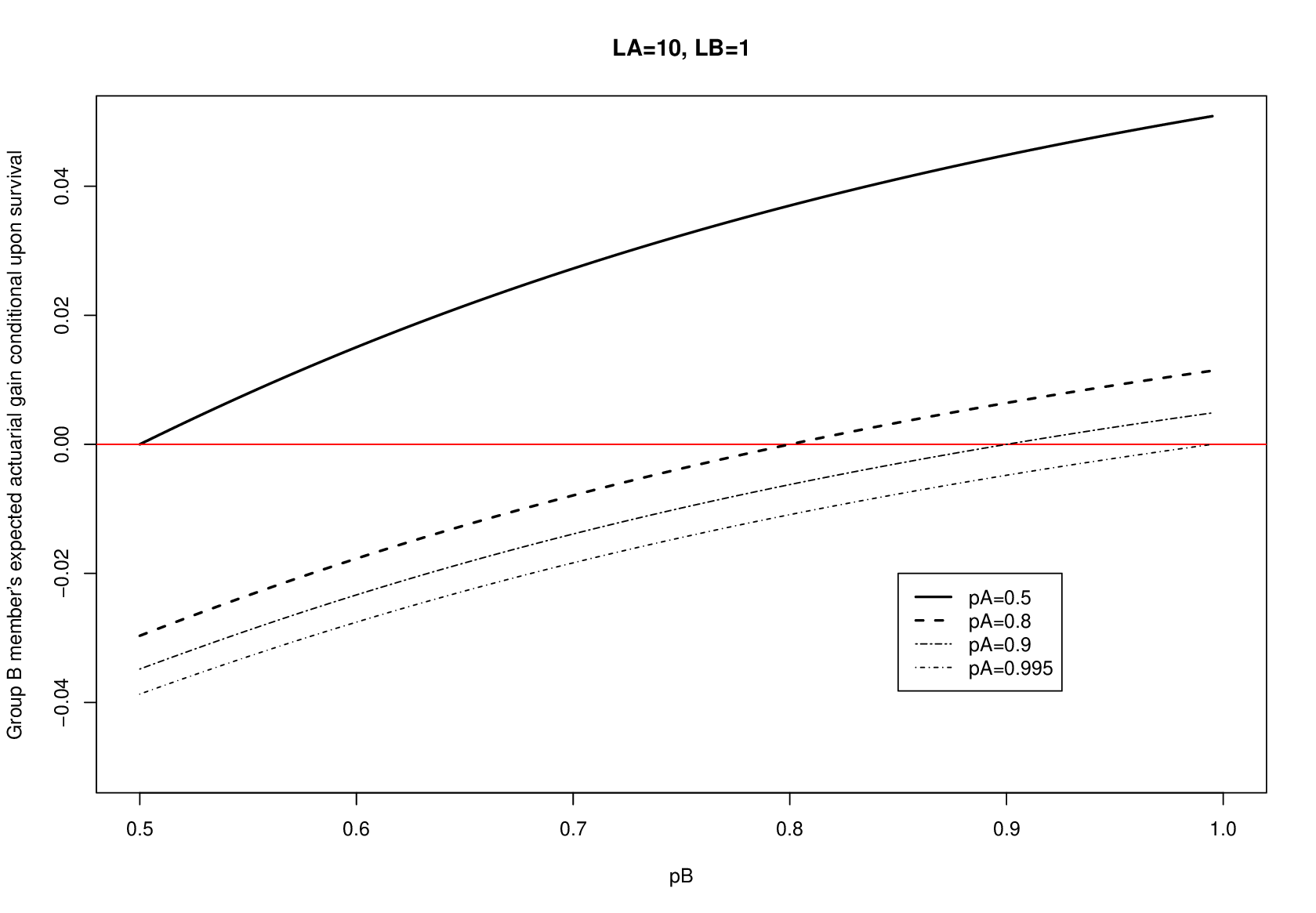}
   }
	 \subfigure[100 members in Group A, 1 member in Group B.]{
  \includegraphics[height=60mm,width=60mm, angle=270]{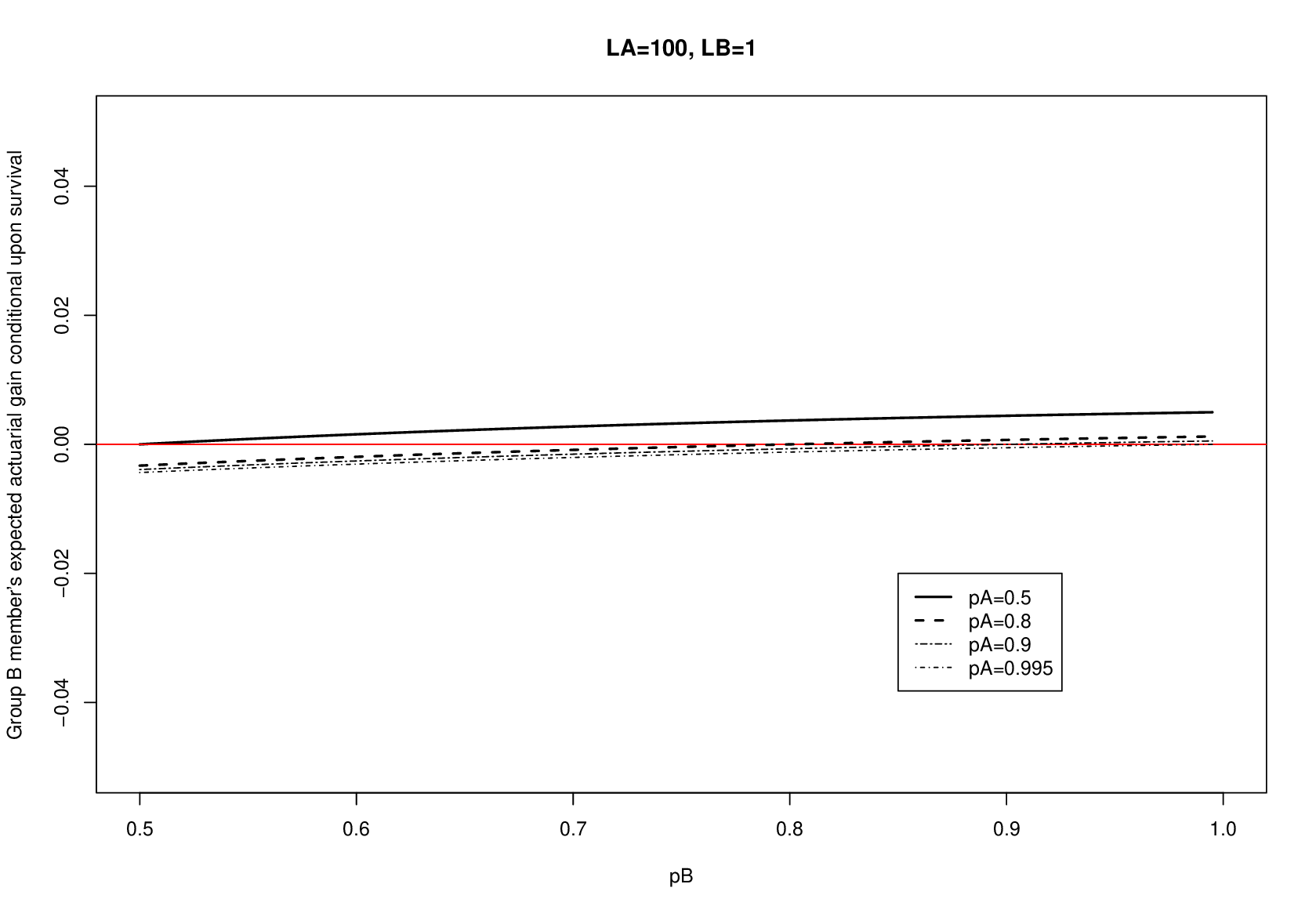}
   }
 \caption{The relative expected actuarial gain, conditional upon survival to time 1, per unit initial contribution for a Group $B$ member as the survival probability $\probsurviveB$ varies, in a heterogeneous \GSAabbrev.  The figures show the gain for a fixed number of members in each group, fixed survival probabilities $\probsurviveA$ and the same initial contribution of 1 unit from each Group $A$ and Group $B$ member.}
 \label{FIGactgainssurvcond}
\end{figure}

\begin{figure}[ht]
 \centering
 \subfigure[10 members in each group.]{
  \includegraphics[height=60mm,width=60mm, angle=270]{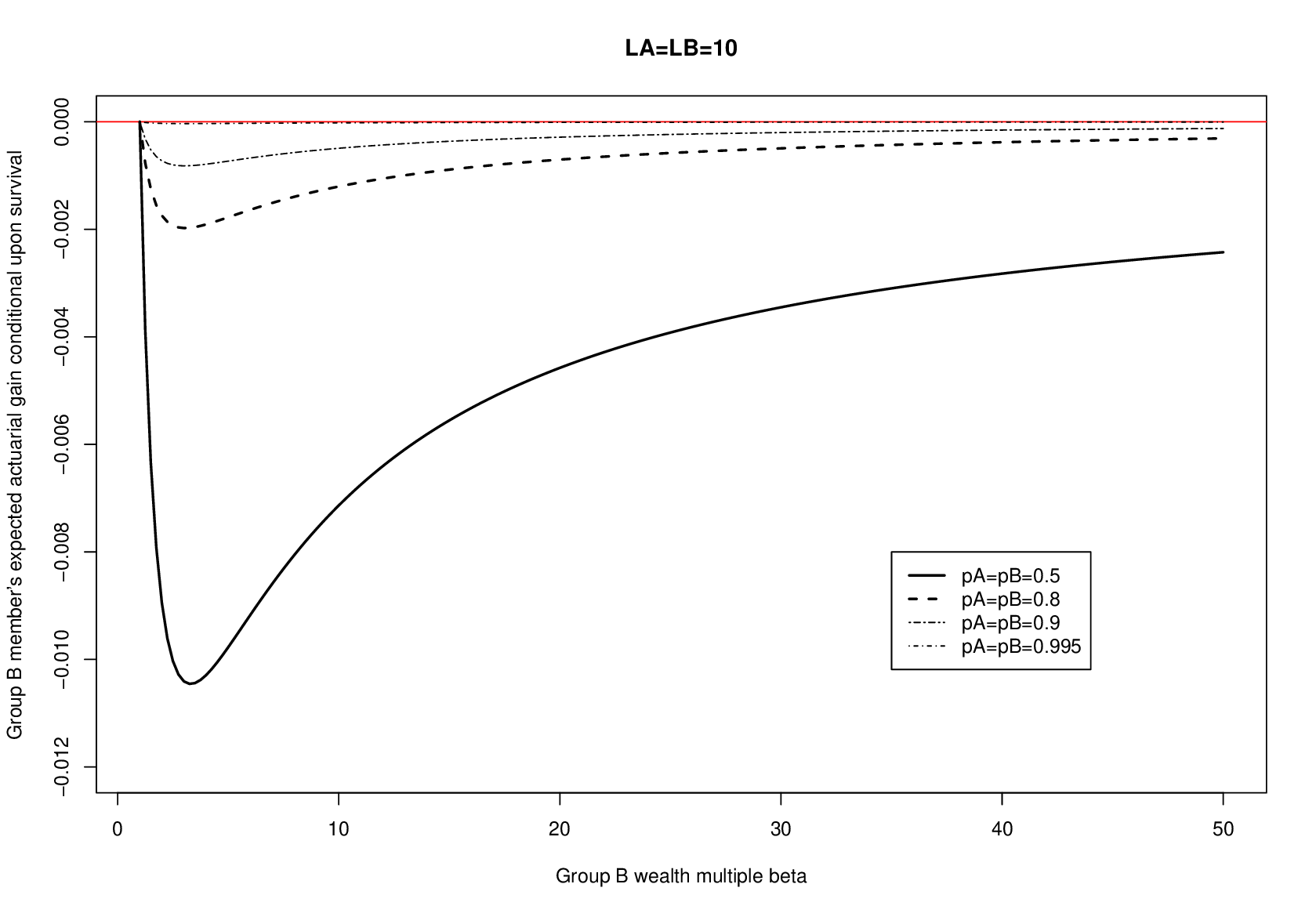}
   }
 \subfigure[100 members in each group.]{
  \includegraphics[height=60mm,width=60mm, angle=270]{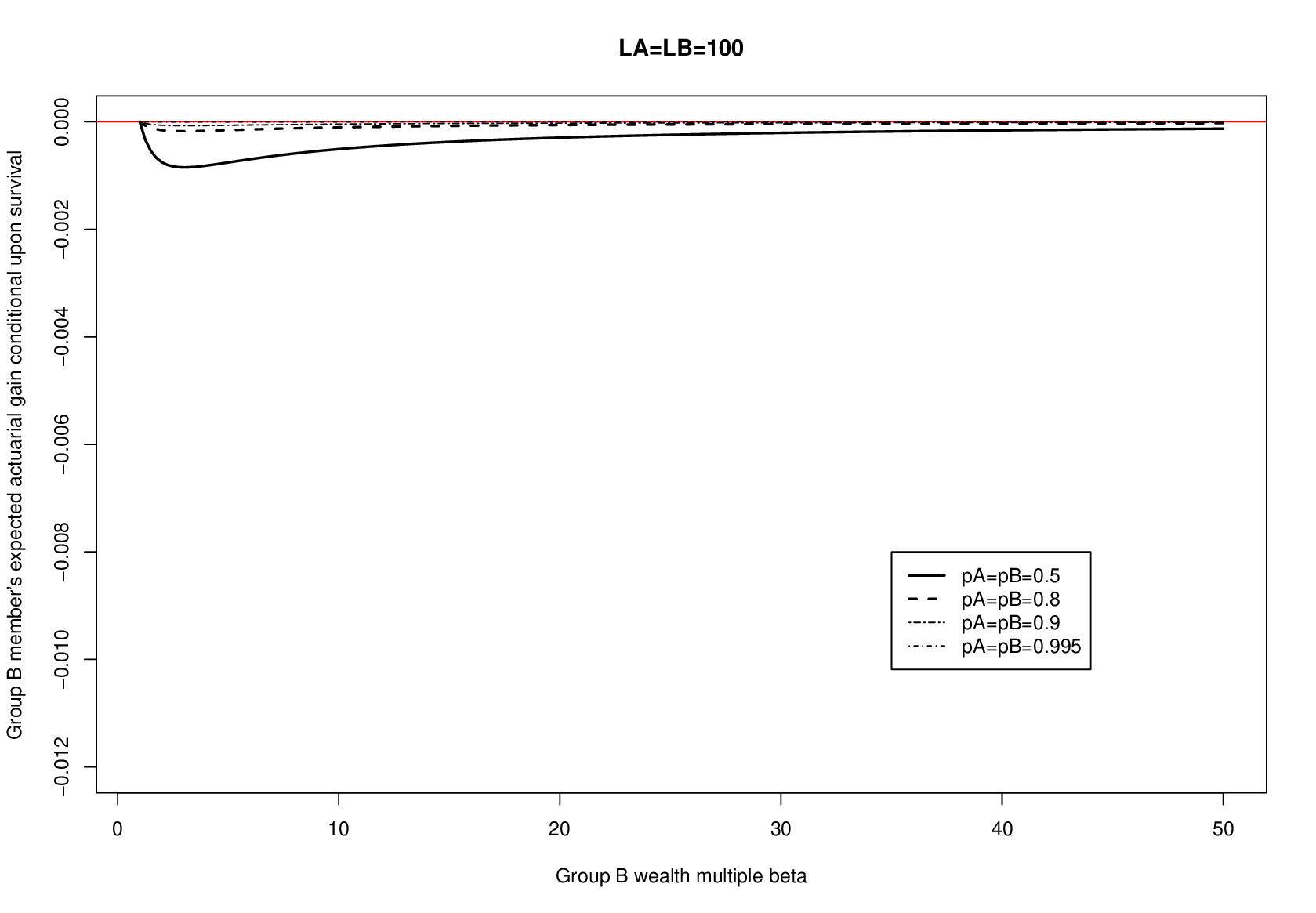}
   }
 \subfigure[10 members in Group A, 1 member in Group B.]{
  \includegraphics[height=60mm,width=60mm, angle=270]{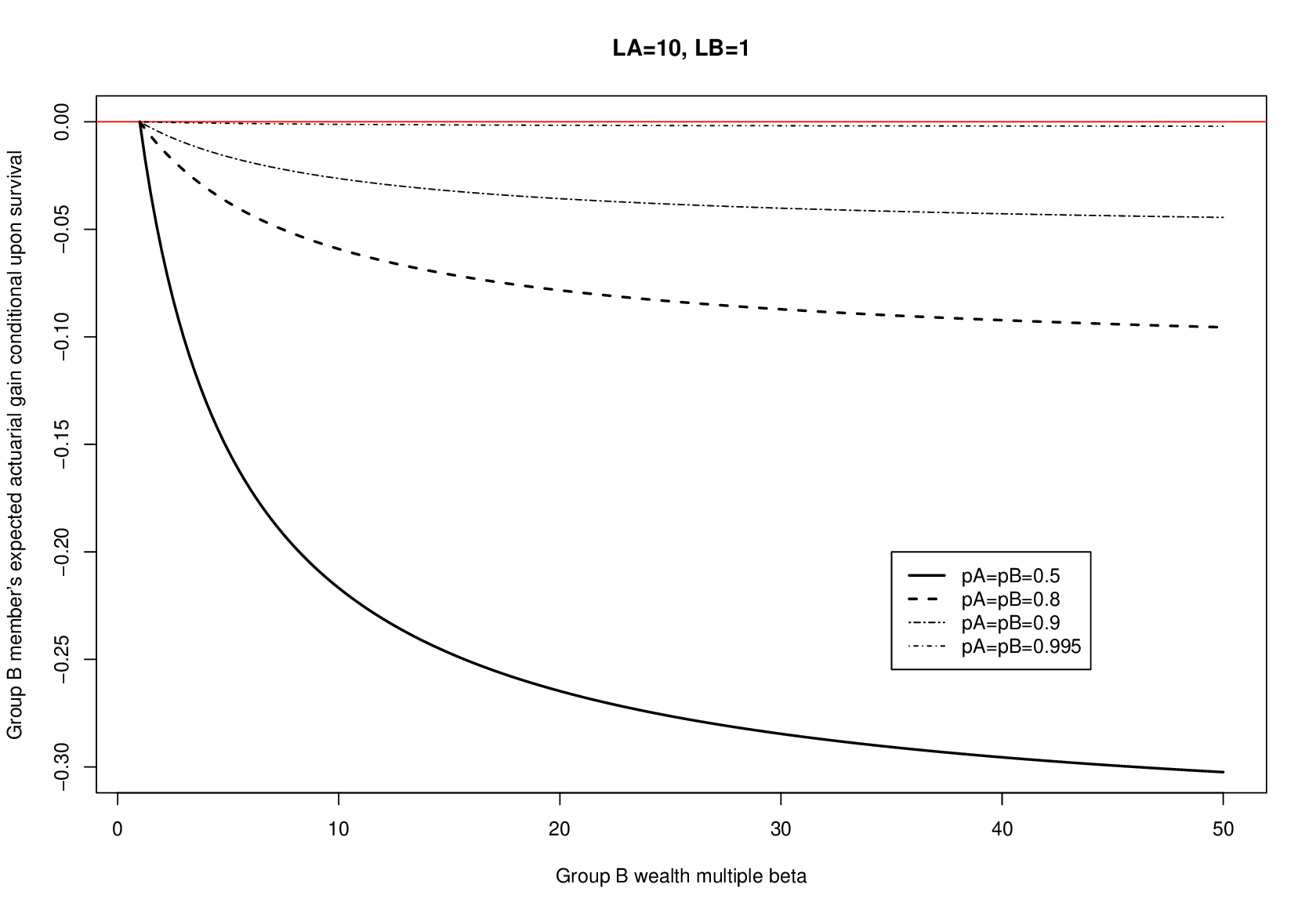}
   }
	 \subfigure[100 members in Group A, 1 member in Group B.]{
  \includegraphics[height=60mm,width=60mm, angle=270]{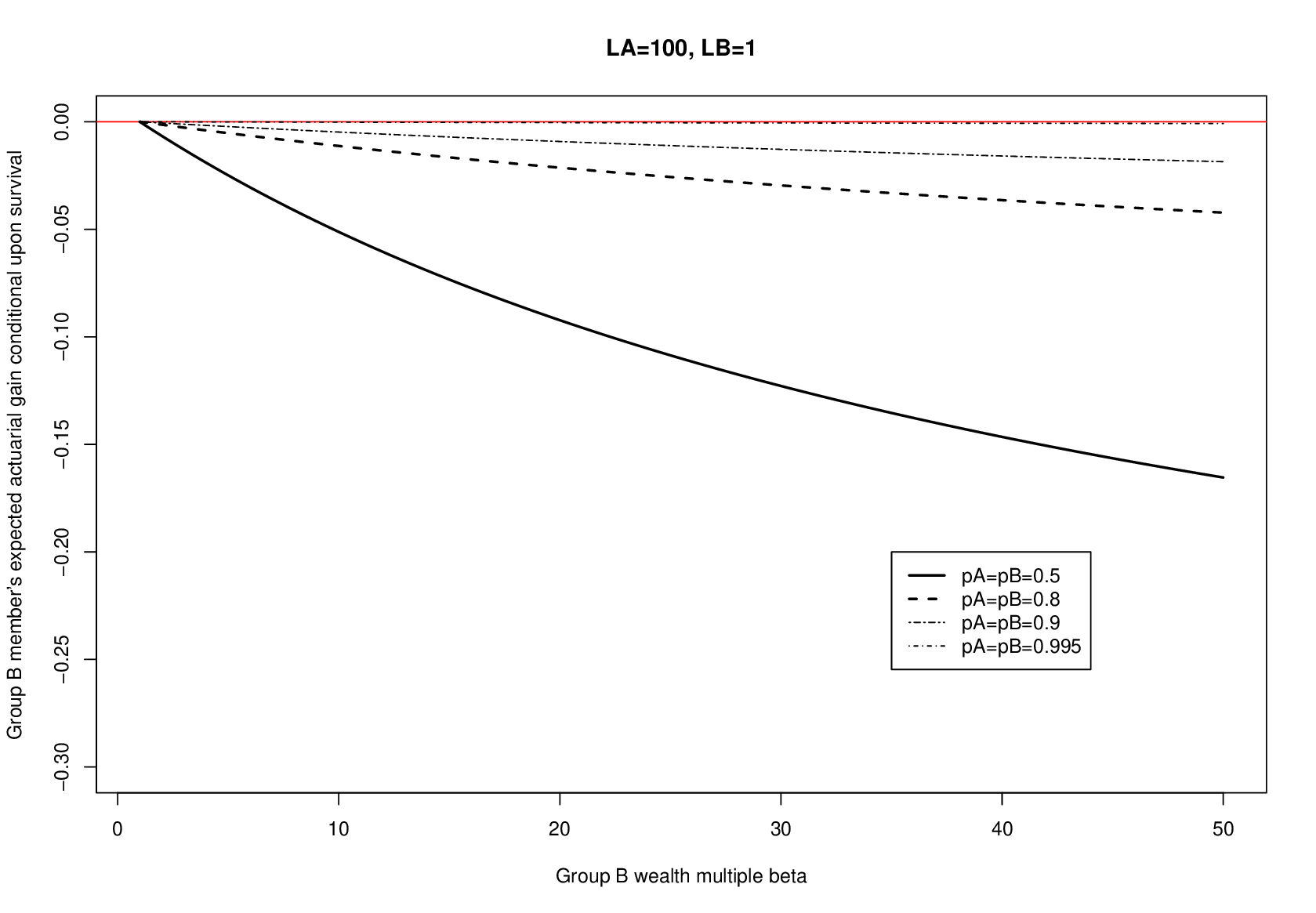}
   }
 \caption{The relative expected actuarial gain, conditional upon survival to time 1, per unit initial contribution for a Group $B$ member as the initial contribution of the Group $B$ member varies, in a heterogeneous \GSAabbrev.  The figures show the gain for a fixed number of members in each group, fixed survival probabilities $\probsurviveA$ and $\probsurviveB$, and initial contribution of 1 unit from each Group $A$ member.}
 \label{FIGactgainscontcond}
\end{figure}

Specifically, we calculate the excess of \namembrB 's expected conditional consumption
\[
\expectation (\consumBzero + \consumBone \, \big\vert \, \textrm{\namembrBspace survives to time 1})
\]
from a heterogeneous \GSAabbrevspace over his expected conditional consumption from a homogeneous \GSAabbrev, in which all members make the same initial contribution as \namembrB.  This gives \namembrB 's relative expected actuarial gain conditional upon survival to time 1.   We divide the relative conditional expected actuarial gain by \namembrB 's initial contribution.  Figures \ref{FIGactgainssurvcond} and \ref{FIGactgainscontcond} illustrate the results.

Again we see that it is only with small or highly heterogeneous groups that the relative conditional gain is significant.  For example, for $\probsurviveA=\probsurviveB=0.9$, \namembrBspace the only member of Group $B$ and and \namembrBspace contributing 10 times the initial amount of each Group $A$ member, \namembrB 's expected conditional actuarial gain is $-2.6\%$ with 10 members in Group $A$, declining to $-0.5\%$ with 100 members in Group $A$.

\subsection{Discussion of the \GSAabbrevspace compared to a \AOFabbrevspace for a heterogeneous group} \label{SUBSECgsaaofhetero}

A mortality risk-sharing arrangement does not create money for a group.  It is simply a rule to allocate the group's funds among the participants.  However, the \AOFabbrevspace and the \GSAabbrevspace approach the sharing of mortality risk differently.  They have different aims.  In the \GSAabbrev, the approach is a group risk-sharing perspective and in the \AOFabbrevspace it is an individual member perspective.

The \GSAabbrevspace assumes that the actual number of deaths in the scheme should be the expected number of deaths, assuming no longevity risk, since this is the basis of the calculation of the future consumption payments.  This is reasonable when there are enough members to sufficiently pool mortality risk, by the Law of Large Numbers.  In the \GSAabbrev, deviations of the observed number of deaths from the expected number of deaths result in adjustments to the consumption payments.  For example, if deaths turn out exactly as expected for Group $A$ but not as expected for Group $B$, then both Group $A$ and $B$ members will have their consumption payments adjusted by the same percentage.

The \AOFabbrevspace starts with the requirement that the fund is actuarially fair for each member, by taking into account the actual scheme membership profile in the calculation of the actuarial gains due to mortality risk-sharing.  Actuarial gains due to deaths in the \AOFabbrevspace result in members' fund values changing by different percentages, according to their individual mortality rate and wealth, rather than by the same percentage as in the \GSAabbrev.

The structure of the \GSAabbrevspace is focused on paying its members a consumption stream.  Members pay an initial contribution and do not individually own a share of the assets of the \GSAabbrev.  This means that individuals give up their claim on their initial contribution in exchange for the consumption stream.

In contrast, the \AOFabbrevspace emphasizes an investment frame, in which the result of mortality risk-sharing is an additional, nonnegative return on an individual's assets.  Participants in an \AOFabbrevspace retain investment control over their own assets.  Indeed, theoretically these assets could include the individual's house which would only need to be sold upon the death of the individual to order to share its financial value among the \AOFabbrevspace members; in the meantime the individual could continue living in their house.  This is unlike the \GSAabbrevspace which, as it pays an income stream to its members, could not have its participants continuing to own and live in their own houses.  The \GSAabbrevspace requires the initial contributions upfront, whereas the \AOFabbrevspace requires it upon the death of the member.

As the mortality risk-sharing rule in the \AOFabbrevspace results in a payment to the estate of the recently deceased, there is less money to be shared among the survivors.  Thus we expect that the survivors in an \GSAabbrevspace should, considered as a group, gain more money than the survivors in the \AOFabbrevspace no matter which future state of the world occurs.

However, this does not hold when we consider each survivor individually in a \GSAabbrev.  Consider the notional fund values in a \GSAabbrev.  Since
\[
\notfundX_{0+} = \notfundX_{0} - \consumXzero = \probsurviveX \consumXzero,
\]
we find
\[
\notfundX_{1} = \fundfraction \cdot \consumXzero = \fundfraction \cdot \frac{\notfundX_{0+}}{\probsurviveX},
\]
which implies
\[
\notfundX_{1} \left\{ \begin{array}{ll}
> \notfundX_{0+} & \textrm{if $\fundfraction > \probsurviveX$} \\
= \notfundX_{0+} & \textrm{if $\fundfraction = \probsurviveX$} \\
< \notfundX_{0+} & \textrm{if $\fundfraction < \probsurviveX$},
\end{array} \right.
\]
for $X \in \{ A,B\}$.  Hence it is possible that the notional fund value of a survivor in Group $A$ increases at time 1 while the notional fund value of a survivor in Group $B$ decreases at time 1.  For example, consider a \GSAabbrevspace with group wealth-mortality characteristics
\[
\grpAchars = (\numalive, \probsurviveA, 1) \quad \textrm{and} \quad \grpBchars = (\numalive, \probsurviveB, 1),
\]
with $\probsurviveB > \probsurviveA$.  Then on the event $[\numdead^{A} = \numdead^{B} = 0]$, the mortality experience adjustment factor defined by equation (\ref{EQNmea}) becomes
\[
\fundfraction = \frac{2 \probsurviveA \probsurviveB}{\probsurviveA + \probsurviveB}.
\]
As $\probsurviveB > \probsurviveA$, it is straightforward to show that
\[
\fundfraction > \probsurviveA \qquad \textrm{and} \qquad \fundfraction < \probsurviveB.
\]
Thus with strictly positive probability,
\[
\notfundA_{1} > \notfundA_{0+} \qquad \textrm{and} \qquad \notfundB_{1} < \notfundB_{0+}, \qquad \textrm{on $[\numdead^{A} = \numdead^{B} = 0]$}.
\]
The interpretation is that when no-one in each group dies, the survivors in Group $B$ lose financially from pooling mortality in the \GSAabbrev.  In comparison, survivors in either group in the corresponding heterogeneous \AOFabbrevspace never lose financially from pooling mortality (see \citealt{donnellyetal13.article}).  In the heterogeneous \AOFabbrevspace when no-one dies in either group by time 1, the fund values at time 1 of each member would equal their fund values at time 0 (ignoring investment returns).

Both structures mean that longevity risk is shared among the group.  We have not analyzed the impact of longevity risk here, but instead refer the reader to \citet{qiaosherris12.article} for the \GSAabbrev.  A related paper is \citet{hanewaldal13.article}.

In conclusion, the attractiveness of the \GSAabbrevspace over the \AOFabbrevspace depends on an individual's circumstances.  One person may prefer to have an income stream paid from the \GSAabbrevspace without having the worry of managing their own assets, another may prefer the flexibility of the \AOFabbrev.  We simply highlight that the structures are different in how they share mortality risk and that this has implications, both financial and for which assets can be used for the pooling.

\section{Summary}
Various risk-sharing pension schemes have been proposed in the literature, which do not involve a sponsoring employer or insurance company.  We have highlighted the connections and differences between some of these schemes.

We have also examined if the \GSAspace is actuarially fair, and we have found that it is not fair for a finite, heterogeneous group.  The primary reason for this result is that the Law of Large Numbers can only be approximately true for a finite group, while the calculation of the consumption payments in the \GSAspace assumes that mortality risk is fully diversified.  The greater the heterogeneity in the fund, the greater is the number of members required to make the assumption of full diversification of mortality risk a reasonable approximation.  Consequently, certain classes of members may be financially disadvantaged in a \GSA.  While it may have been the intention of the authors who proposed the \GSAspace that they are only operated for large numbers of people, this is not often clearly stated.   

The relative attractiveness of the examined schemes for different types of individuals is an important question that we leave for future work.

\section*{Acknowledgements}
The author gratefully acknowledges the financial support of the 2013 Individual Grant received from The Actuarial Foundation.  The author thanks two anonymous referees whose thoughtful comments helped to improve the paper.

\newpage
\appendix

\section{Proofs}

\samesize*

%
%

\begin{proof}
In a heterogeneous \GSAabbrevspace in which the wealth-mortality characteristics of
\begin{itemize}
	\item Group $A$ are $\grpAchars = (\numalive, \probsurvive, 1)$, i.e. $\notfundA_{0} = 1$, and
	\item Group $B$ are $\grpBchars = (\numalive, \probsurvive, \wealthrich)$, i.e. $\notfundB_{0} = \wealthrich$,
\end{itemize}
denote by
\begin{itemize}
	\item $\consumBzero (\wealthrich)$ the consumption at time 0 of a member of Group $B$,
	\item $\consumBone (\wealthrich)$ the consumption at time 1 of a surviving member of Group $B$,
	\item $\fnbobsurvives (\wealthrich) := \expectation (\consumBzero (\wealthrich) + \consumBone(\wealthrich) \, \big\vert \, \textrm{\namembrBspace survives to time 1})$ the expected consumption of \namembrBspace conditional upon his survival to time 1, and
	\item $\fnbobeither (\wealthrich) := \expectation (\consumBzero(\wealthrich) + \consumBone(\wealthrich))$ the expected consumption of \namembrB.
\end{itemize}

We begin by showing that, conditional upon his survival to time 1, \namembrB 's expected consumption in the heterogeneous \GSAabbrevspace is different than in a homogeneous \GSAabbrev, in which all surviving members have the same amount of wealth as \namembrB.  Specifically, we prove that
\begin{equation} \label{INEQUALbobsurvives}
\fnbobsurvives (\wealthrich)
 \left\{ \begin{array}{ll}
< \wealthrich \fnbobsurvives (1) & \textrm{if $\wealthrich > 1$} \\
= \wealthrich \fnbobsurvives (1) & \textrm{if $\wealthrich = 1$} \\
> \wealthrich \fnbobsurvives (1) & \textrm{if $\wealthrich < 1$},
\end{array} \right.
\end{equation}
in which $\wealthrich \fnbobsurvives (1)$ is \namembrB 's conditional expected consumption in a homogeneous \GSAabbrevspace in which all members make the same initial contribution $\wealthrich$.

From (\ref{EQNgsaconsum}),
\[
\notfundX_{0+} := \notfundX_{0} - \consumXzero = \notfundX_{0} \left( 1 - \frac{1}{\ddot{a}_{0}^{X}} \right) = \notfundX_{0} \frac{\probsurvive}{1+\probsurvive},
\]
so that from (\ref{EQNmea}) and (\ref{EQNhetgsaconsumX}) we find
\[
\begin{split}
\expectation ( \consumBone(\wealthrich) \, \vert \, \textrm{\namembrBspace survives to time 1}) = & \expectation ( \consumXzero \cdot \fundfraction \, \vert \, \textrm{\namembrBspace survives to time 1})\\
= & \expectation \left( \frac{\wealthrich}{1 + \probsurvive} \cdot \frac{\probsurvive (\wealthrich + 1) \numalive}{\numalive - \numdead^{A} + \wealthrich (\numalive - \numdead^{B})} \, \bigg\vert \, \textrm{\namembrBspace survives to time 1} \right).
\end{split}
\]
Set $\probdie := 1 -\probsurvive$.  Since $\numdead^{A} \sim BIN(\numalive, \probdie)$ and $\numdead^{B} \sim BIN(\numalive - 1, \probdie)$, we get
\[
\fnbobsurvives (\wealthrich) = \frac{\wealthrich}{1 + \probsurvive} + \frac{\probsurvive}{1 + \probsurvive} \sum_{n=0}^{\numalive} \sum_{m=0}^{\numalive - 1} \frac{\wealthrich (\wealthrich+1)}{\numalive - n + \wealthrich (\numalive - m)} \, \binom{\numalive}{n} \binom{\numalive - 1}{m} \probdie^{n+m} \probsurvive^{2 \numalive - 1 - (n+m)}.
\]
Define
\[
\fnsummand(n,m,\wealthrich) := \frac{\wealthrich (\wealthrich+1)}{\numalive - n + \wealthrich (\numalive - m)} \, \binom{\numalive}{n} \binom{\numalive - 1}{m} \probdie^{n+m} \probsurvive^{2 \numalive - 1 - (n+m)},
\]
which allows us to write more compactly
\[
\begin{split}
\fnbobsurvives (\wealthrich) & = \frac{\wealthrich}{1 + \probsurvive} + \frac{\probsurvive}{1 + \probsurvive} \sum_{n=0}^{\numalive} \sum_{m=0}^{\numalive - 1} \fnsummand(n,m,\wealthrich) \\
& = \frac{\wealthrich}{1 + \probsurvive} + \frac{\probsurvive}{1 + \probsurvive} \sum_{n=0}^{\numalive - 1} \sum_{m=0}^{\numalive - 1} \fnsummand(n,m,\wealthrich) + \frac{\probsurvive}{1 + \probsurvive} \sum_{m=0}^{\numalive - 1} \fnsummand(\numalive,m,\wealthrich).
\end{split}
\]

First, as
\[
\sum_{n=0}^{\numalive - 1} \sum_{m=0}^{\numalive - 1} \fnsummand(n,m,\wealthrich) = \sum_{n=0}^{\numalive - 1} \fnsummand(n,n,\wealthrich) + \indicatorfn_{\left[ \numalive \geq 2 \right]} \sum_{n=0}^{\numalive - 2} \sum_{m=n+1}^{\numalive - 1}  \left[ \fnsummand(n,m,\wealthrich) + \fnsummand(m,n,\wealthrich) \right]
\]
and
\[
\sum_{m=0}^{\numalive - 1} h(\numalive,m,\wealthrich) = (\wealthrich+1) \frac{\probdie^{\numalive} (1 - \probdie^{\numalive})}{\probsurvive},
\]
then
\[
\begin{split}
 \fnbobsurvives (\wealthrich) & =  \frac{\wealthrich}{1 + \probsurvive} + \frac{\probsurvive}{1 + \probsurvive} \left( \sum_{n=0}^{\numalive - 1} \fnsummand(n,n,\wealthrich) + \indicatorfn_{\left[ \numalive \geq 2 \right]} \sum_{n=0}^{\numalive - 2} \sum_{m=n+1}^{\numalive - 1} \left[ \fnsummand(n,m,\wealthrich) + \fnsummand(m,n,\wealthrich) \right] \right) \\
& + \frac{(\wealthrich+1) \probdie^{\numalive} (1 - \probdie^{\numalive})}{1 + \probsurvive}.
\end{split}
\]

Thus
\begin{equation} \label{EQNintermedactgains}
\begin{split}
\wealthrich \fnbobsurvives (1) - \fnbobsurvives (\wealthrich) & = \frac{\probsurvive}{1 + \probsurvive} \sum_{n=0}^{\numalive - 1} \left[ \wealthrich \fnsummand \left( n,n,1 \right) - \fnsummand(n,n,\wealthrich) \right] \\
& + \indicatorfn_{\left[ \numalive \geq 2 \right]} \frac{\probsurvive}{1 + \probsurvive} \sum_{n=0}^{\numalive - 2} \sum_{m=n+1}^{\numalive - 1} \left( \wealthrich \left[ \fnsummand(n,m,1) + \fnsummand(m,n,1) \right] - \left[ \fnsummand(n,m,\wealthrich) + \fnsummand(m,n,\wealthrich) \right] \right) \\
& + \frac{(\wealthrich - 1) \probdie^{\numalive} (1 - \probdie^{\numalive})}{1 + \probsurvive}.
\end{split}
\end{equation}

Next, after some algebra, we find for $n=0,1,\ldots,\numalive - 1$ that
\[
\fnsummand\left(n,n,\wealthrich \right) = \wealthrich \, \fnsummand \left( n,n,1 \right)
\]
and
\[
\begin{split}
& \fnsummand(n,m,\wealthrich) + \fnsummand(n,m,\wealthrich) \\
= & \frac{\wealthrich (\wealthrich + 1)}{\numalive} \binom{\numalive}{n} \binom{\numalive}{m} \probdie^{n+m} \probsurvive^{2 \numalive - 1 - (n+m)} \left[ \frac{\numalive - m}{\numalive - n + \wealthrich (\numalive - m)} + \frac{\numalive - n}{\numalive - m + \wealthrich (\numalive - n)} \right],
\end{split}
\]
which allows us to simplify equation (\ref{EQNintermedactgains}) to
\[
\begin{split}
& \wealthrich \fnbobsurvives (1) - \fnbobsurvives (\wealthrich) \\
= & \frac{\wealthrich - 1}{1 + \probsurvive} \\
& \cdot \left[ \frac{\wealthrich \probsurvive}{\numalive} \indicatorfn_{\left[ \numalive \geq 2 \right]} \sum_{n=0}^{\numalive - 2} \sum_{m=n+1}^{\numalive - 1} \binom{\numalive}{n} \binom{\numalive}{m}  \frac{(n-m)^{2} \probdie^{n+m} \probsurvive^{2 \numalive - 1 - (n+m)}}{\left[ \numalive - n + \wealthrich (\numalive - m) \right] \left[ \numalive - m + \wealthrich (\numalive - n) \right]} + \probdie^{\numalive} (1 - \probdie^{\numalive}) \right].
\end{split}
\]
From the positivity of the terms in the square bracket we obtain the required inequality (\ref{INEQUALbobsurvives}). 

Now turning to \namembrB 's expected consumption, we show that his expected consumption in the heterogeneous \GSAabbrevspace is different than in a homogeneous \GSAabbrev, in which all surviving members have the same amount of wealth as \namembrB.  Recall that his expected consumption is given by
\[
\fnbobeither (\wealthrich) = \expectation (\consumBzero(\wealthrich) + \consumBone(\wealthrich)).
\]
We prove that
\begin{equation} \label{INEQUALbobeither}
\fnbobeither (\wealthrich)
 \left\{ \begin{array}{ll}
< \wealthrich  & \textrm{if $\wealthrich > 1$} \\
= \wealthrich  & \textrm{if $\wealthrich = 1$} \\
> \wealthrich & \textrm{if $\wealthrich < 1$}.
\end{array} \right.
\end{equation}
We assume that if everyone in the \GSAabbrevspace dies by time 1 then the estates of the members receive their individual notional fund values.  Then
\[
\begin{split}
\fnbobeither (\wealthrich) & = \probsurvive \fnbobsurvives (\wealthrich) + \probdie \expectation ( \consumBzero (\wealthrich)  + \consumBone(\wealthrich) \, \vert \, \textrm{\namembrBspace dies by time 1}) \\
& = \probsurvive \fnbobsurvives (\wealthrich) + \probdie \left( \frac{\wealthrich}{1 + \probsurvive} + \frac{\wealthrich \probsurvive}{1 + \probsurvive} \probdie^{2 \numalive - 1} \right),
\end{split}
\]
from which we obtain
\[
\wealthrich \fnbobeither (1) - \fnbobeither (\wealthrich) = \probsurvive \left( \wealthrich \fnbobsurvives (1) - \fnbobsurvives (\wealthrich) \right).
\]
Upon noting that the homogeneous \GSAabbrevspace is actuarially fair, i.e. $\fnbobeither (1) = 1$, and that $\probsurvive \in (0,1)$, the inequality (\ref{INEQUALbobeither}) follows from (\ref{INEQUALbobsurvives}).
\end{proof}

%
%
%
%

\bobonly*

\begin{proof}
In a heterogeneous \GSAabbrevspace in which the wealth-mortality characteristics of
\begin{itemize}
	\item Group $A$ are $\grpAchars = (\numalive^{A}, \probsurvive, 1)$, i.e. $\notfundA_{0} = 1$, and
	\item Group $B$ are $\grpBchars = (1, \probsurvive, \wealthrich)$, i.e. $\notfundB_{0} = \wealthrich$,
\end{itemize}
denote by
\begin{itemize}
	\item $\consumBzero (\wealthrich)$ the consumption at time 0 of \namembrB,
	\item $\consumBone (\wealthrich)$ the consumption at time 1 of \namembrBspace if he survives to time 1,
	\item $\fnbobsurvives (\wealthrich) := \expectation (\consumBzero (\wealthrich) + \consumBone(\wealthrich) \, \big\vert \, \textrm{\namembrBspace survives to time 1})$ the expected consumption of \namembrBspace conditional upon his survival to time 1, and
	\item $\fnbobeither (\wealthrich) := \expectation (\consumBzero(\wealthrich) + \consumBone(\wealthrich))$ the expected consumption of \namembrB.
\end{itemize}

We begin by showing that, conditional upon his survival to time 1, \namembrB 's expected consumption in the heterogeneous \GSAabbrevspace is different than in a homogeneous \GSAabbrev, in which all surviving members have the same amount of wealth as \namembrB.  As before, we show that
\begin{equation} \label{INEQUALbobsurvivesii}
\fnbobsurvives (\wealthrich)
 \left\{ \begin{array}{ll}
< \wealthrich \fnbobsurvives (1) & \textrm{if $\wealthrich > 1$} \\
= \wealthrich \fnbobsurvives (1) & \textrm{if $\wealthrich = 1$} \\
> \wealthrich \fnbobsurvives (1) & \textrm{if $\wealthrich < 1$}.
\end{array} \right.
\end{equation}
From (\ref{EQNmea}) and (\ref{EQNhetgsaconsumX}),
\[
\begin{split}
& \expectation ( \consumBone(\wealthrich) \, \vert \, \textrm{\namembrBspace survives to time 1}) \\
= & \expectation \left( \frac{\wealthrich}{1 + \probsurvive} \cdot \frac{\probsurvive \left( \numalive + \wealthrich \right)}{\numalive - \numdead^{A} + \wealthrich} \, \bigg\vert \, \textrm{\namembrBspace survives to time 1} \right).
\end{split}
\]
Set $\probdie := 1 -\probsurvive$.  Since $\numdead^{A} \sim BIN(\numalive, \probdie)$, we get
\[
\fnbobsurvives (\wealthrich) = \frac{\wealthrich}{1 + \probsurvive} + \frac{\probsurvive}{1 + \probsurvive} \sum_{n=0}^{\numalive} \frac{\wealthrich (\numalive + \wealthrich)}{\numalive + \wealthrich - n} \, \binom{\numalive}{n} \probdie^{n} \probsurvive^{\numalive - n}.
\]

Thus
\[
\begin{split}
\wealthrich \fnbobsurvives (1) - \fnbobsurvives (\wealthrich) & = \frac{\wealthrich \probsurvive}{1 + \probsurvive} \sum_{n=0}^{\numalive}  \binom{\numalive}{n} \probdie^{n} \probsurvive^{\numalive - n} \left[ \frac{\numalive + 1}{\numalive + 1 - n} - \frac{\numalive + \wealthrich}{\numalive + \wealthrich - n} \right] \\
& = \frac{\wealthrich (\wealthrich - 1) \probsurvive}{1 + \probsurvive} \sum_{n=0}^{\numalive} \binom{\numalive}{n} \probdie^{n} \probsurvive^{\numalive - n} \frac{n}{(\numalive + 1 - n) (\numalive + \wealthrich - n)},
\end{split}
\]
so that the sign of $\wealthrich \fnbobsurvives (1) - \fnbobsurvives (\wealthrich)$ is inherited from the sign of $\wealthrich - 1$.  Thus we obtain the inequality (\ref{INEQUALbobsurvivesii}).

Now turning to \namembrB 's overall expected consumption, the proof that
\[
\fnbobeither (\wealthrich)
 \left\{ \begin{array}{ll}
< \wealthrich  & \textrm{if $\wealthrich > 1$} \\
= \wealthrich  & \textrm{if $\wealthrich = 1$} \\
> \wealthrich & \textrm{if $\wealthrich < 1$}
\end{array} \right.
\]
is identical to that in Proposition \ref{PROPNsamesizesgsa}.
\end{proof}

\newpage

\section{Figures}

\begin{thebibliography}{9}
\providecommand{\natexlab}[1]{#1}
\providecommand{\url}[1]{\texttt{#1}}
\expandafter\ifx\csname urlstyle\endcsname\relax
  \providecommand{\doi}[1]{doi: #1}\else
  \providecommand{\doi}{doi: \begingroup \urlstyle{rm}\Url}\fi

\bibitem[Donnelly(2014)]{donnelly11.article}
C.~Donnelly.
\newblock Quantifying mortality risk in small defined-benefit pension schemes.
\newblock \emph{Scandinavian Actuarial Journal}, 2014\penalty0 (1):\penalty0
  41--57, 2014.

\bibitem[Donnelly et~al.(2013)Donnelly, Guill{\'{e}}n, and
  Nielsen]{donnellyetal13.article}
C.~Donnelly, M.~Guill{\'{e}}n, and J.P. Nielsen.
\newblock Exchanging uncertain mortality for a cost.
\newblock \emph{Insurance: Mathematics and Economics}, 52\penalty0
  (1):\penalty0 65--76, 2013.

\bibitem[Donnelly et~al.(2014)Donnelly, Guill{\'{e}}n, and
  Nielsen]{donnellyetal14.article}
C.~Donnelly, M.~Guill{\'{e}}n, and J.P. Nielsen.
\newblock Bringing cost transparency to the life annuity market.
\newblock \emph{Insurance: Mathematics and Economics}, 56:\penalty0 14--27,
  2014.

\bibitem[Hanewald et~al.(2013)Hanewald, Piggott, and
  Sherris]{hanewaldal13.article}
K.~Hanewald, J.~Piggott, and M.~Sherris.
\newblock Individual post-retirement longevity risk management under systematic
  mortality risk.
\newblock \emph{Insurance: Mathematics and Economics}, 52\penalty0
  (1):\penalty0 87--97, 2013.

\bibitem[Piggott et~al.(2005)Piggott, Valdez, and
  Detzel]{piggottetal05.article}
J.~Piggott, E.A. Valdez, and B.~Detzel.
\newblock The simple analytics of a pooled annuity fund.
\newblock \emph{Journal of Risk and Insurance}, 72\penalty0 (3):\penalty0
  497--520, 2005.

\bibitem[Pension Protection Fund(2013)]{PPF2013.misc}
PPF.
\newblock The purple book: {DB} pensions universe risk profile, 2013.
\newblock Produced by the Pension Protection Fund and The Pensions Regulator.
  Published on 5 November 2013.
  \url{http://www.pensionprotectionfund.org.uk/Pages/ThePurpleBook.aspx}.

\bibitem[Qiao and Sherris(2013)]{qiaosherris12.article}
C.~Qiao and M.~Sherris.
\newblock Managing systematic mortality risk with group self pooling and
  annuitization schemes.
\newblock \emph{Journal of Risk and Insurance}, 80\penalty0 (4):\penalty0
  949â€“--974, 2013.

\bibitem[Sabin(2010)]{Sabin10.unpublished}
M.J. Sabin.
\newblock Fair tontine annuity.
\newblock \url{http://ssrn.com/abstract=1579932}, March 2010.

\bibitem[Stamos(2008)]{stamos08.article}
M.Z. Stamos.
\newblock Optimal consumption and portfolio choice for pooled annuity funds.
\newblock \emph{Insurance: Mathematics and Economics}, 43\penalty0
  (1):\penalty0 56--68, 2008.

\end{thebibliography}
\end{document}